\documentstyle[pra,aps,amssymb,epsf,floats]{revtex}
\begin{document}
\title{Nullification of the Nullification}
\author{D.M.Appleby\thanks{electronic address: 
D.M.Appleby@qmw.ac.uk}}
\address{Department of Physics, Queen Mary
University of London,  Mile End Rd, London E1 4NS,
UK}
\maketitle
\begin{abstract}
A recent claim by Meyer, Kent and Clifton (MKC), that their models
``nullify'' the Kochen-Specker theorem, has attracted much
comment.  In this paper we present a new
counter-argument, based on the fact that a classical measurement
reveals, not simply a pre-existing value, but  pre-existing 
classical information.  In the MKC models measurements do not
generally reveal pre-existing classical information. 
Consequently, the Kochen-Specker theorem is not nullified.  We go
on to prove a generalized version of the Kochen-Specker theorem,
applying to non-ideal quantum measurements.   The theorem was
inspired by the work of Simon \emph{et al} and Larsson (SBZL). 
However, there is a loophole in SBZL's argument, which means that
their result is invalid (\emph{operational} non-contextuality  is
\emph{not} inconsistent with the empirical predictions of quantum
mechanics).  Our treatment resolves this difficulty.   We conclude
by discussing the question, whether the MKC models can reproduce
the empirical predictions of quantum mechanics.
\end{abstract}
\section{Introduction}
Prior to the work of  Meyer~\cite{Meyer}, the Kochen-Specker
theorem
\cite{Bell1,Koch,MerminB,PeresBk,BubBk} (KS theorem) was
generally  regarded as one of the key foundational results of
quantum mechanics, establishing  one of the most important ways in
which quantum mechanics enforces a radical departure from the
assumptions of classical physics.  Classically, a measurement is a
process which reveals the pre-existing value of the
observable measured.  Until recently it was generally  accepted
that the KS theorem shows that quantum mechanics cannot be
interpreted in such a way as to preserve this  feature of
classical physics.

Unfortunately, the KS theorem, as originally formulated, only
applies to ideal, or von Neumann measurement processes.  In
practice, strict ideality is seldom, and perhaps never actually
attainable.   It follows that the KS theorem, in its original
form, is not sufficient to demonstrate a contradiction between
classical assumptions and the empirically verifiable predictions
of quantum mechanics, regarding the outcome of real, laboratory
measurements.

At first sight, this may not seem a serious
problem.  The natural assumption would be that the original KS
theorem is a limiting case of a more general theorem, which does
apply to non-ideal measurements.  We will eventually argue that
this is, in fact, the correct assumption.  However, the question is
greatly complicated by the work of Meyer~\cite{Meyer},
Kent~\cite{Kent1} and Clifton and Kent~\cite{Kent2} (MKC in the
sequel), who have argued that ``finite precision measurement
nullifies the Kochen-Specker theorem''.  

MKC's claimed ``nullification'' of the KS theorem has given rise
to some controversy.  It has been discussed by (in chronological
order) Cabello~\cite{Cabel}, Havlicek
\emph{et al}~\cite{Havli},
Mermin~\cite{Mermin1}, Appleby\cite{me3,me2}, Simon \emph{et
al}\cite{Zeil1}, Larsson~\cite{Lars}, Simon~\cite{Simon2}, 
Cabello~\cite{Cabel2} and Boyle and Schafir~\cite{Boyle}.  For
further discussion of the problem of demonstrating an empirical
contradiction with the predictions of non-contextual hidden
variables theories see Cabello and
Garc\'{i}a-Alcaine\cite{Cabel3}, 
 Basu \emph{et al}~\cite{Basu}, Cabello~\cite{Cabel4},
Michler~\emph{et al}\cite{Michler}, Simon \emph{et al}~\cite{Zeil3}
and Cereceda~\cite{Cerec}.  Finally, it should be noted that MKC's
argument was inspired by the previous  work of
Pitowsky~\cite{Pitowski},  Hales and Straus~\cite{Hales} and
Godsil and Zaks~\cite{Godsil} (we briefly comment on Pitowsky's
work in the conclusion).

The arguments in this paper are largely new.  They go
significantly beyond any that have previously been given.

We begin by nullifying MKC's claimed nullification.   That is,
we show that MKC's ingenious mathematical constructions, though
deeply interesting in their own right, do not invalidate the
essential physical point of the KS theorem.   Having  cleared the
conceptual ground, we then go on to establish  a
 generalized   version of the KS theorem, which does apply to
non-ideal measurements.  In order to make the treatment
comprehensive we conclude by discussing a number of related
questions.  In particular, we discuss a recent claim by
Cabello~\cite{Cabel2}, that the MKC models do not reproduce the
empirically verifiable predictions of quantum mechanics.

Our main criticism of MKC's argument is contained in
Sections~\ref{sec:WhatObs}--\ref{sec:NullFurther}.

Section~\ref{sec:WhatObs} contains some preliminary
considerations.  MKC assume
that an experimenter can never precisely know what has 
``actually'' been measured.  This assumption plays an important
role in their argument because it enables them to postulate that
it is physically impossible to measure each of the observables in a
KS-uncolourable set.  We show that the assumption depends on some
misconceptions regarding non-ideal quantum measurements.

Section~\ref{sec:ExtendKS1} contains the crux of the critical
part of our  argument.  The physical point of the KS
theorem is to show that quantum measurements cannot generally be
interpreted as measurements in the classical sense.  
A classical measurement is not simply a process which
reveals a pre-existing value. 
 Rather, it is a process which
reveals   a  pre-existing piece of classical
\emph{information}, represented by a proposition of the form
``observable
$A$ took value
$x$''.  For this condition to be satisfied it is essential that 
the value $x$ and observable $A$ \emph{both} be adequately
specified.   It appears to MKC that their models nullify the KS
theorem because, in their models, a measurement does always reveal
the value of \emph{something}.  However, it does not  reveal
the precise identity of that something.   Since
the MKC valuations are radically discontinuous this means that 
measurements do not   generally reveal any classical information. 
Consequently, they cannot generally be regarded as measurements in
the classical sense.  It follows that the KS theorem is not
nullified.

Section~\ref{sec:NullFurther} contains some supplementary
considerations, which reinforce the conclusion reached in
Section~\ref{sec:ExtendKS1}.

The argument just outlined goes beyond the arguments given in our
previous papers because it shows that MKC's claim, to have
nullified the KS theorem, is simply false.  In
Appleby~\cite{me3,me2} we showed that the MKC models are highly
non-classical.  In fact, we showed in ref.~\cite{me2}  that the
MKC models exhibit a novel kind of contextuality, which is even
more strikingly at variance with classical assumptions than the
usual kind of contextuality, featuring in the KS theorem. 
Consequently, the argument in ref.~\cite{me2} is, by itself,
sufficient to refute MKC's suggestion, that their models provide a
classical explanation of non-relativistic quantum mechanics. The
argument in ref.~\cite{me3}, though less clear-cut, lends
additional support to this conclusion.  However, neither argument
directly bears on MKC's claim, to have nullified the KS theorem as
such.   Indeed, at the time we wrote these papers it appeared to
us that MKC actually had ``nullified the KS theorem \emph{strictly 
so-called}'' (as we put it).

It is important to establish that finite precision does
unequivocally \emph{not} nullify the KS theorem because, in the
absence of such a demonstration, it is impossible to achieve an
unobstructed understanding of contextuality, as it applies in a
real, laboratory setting.  The question has immediate, practical
relevance, in view of the  current interest in experimental
investigations of 
contexuality~\cite{Zeil1,Lars,Simon2,Cabel3,Basu,Cabel4,Michler,Zeil3,Cerec}.

Having  cleared the conceptual ground in
Sections~\ref{sec:WhatObs}--\ref{sec:NullFurther} we go on, in
Section~\ref{sec:ExtendKS2}, to formulate and prove a generalized
KS theorem, applying to non-ideal measurements.  The argument in
this section is inspired by the argument of Simon~\emph{et
al}~\cite{Zeil1,Simon2} and Larsson~\cite{Lars} (SBZL in the
sequel).   However, there are some important differences. SBZL
work with an ``operational'' concept of
contextuality.     Their
motive for introducing this concept is to circumvent MKC's
argument.  Clearly, the motive no longer applies, once it
is established that MKC's argument is invalid.   
  We are consequently able to formulate a
generalized KS theorem in terms of the ordinary concept of
contextuality.  This has several advantages.  In the first place
it means that the result we prove is a \emph{straightforward}
generalization of the ordinary KS theorem.  In the second place,
there is a loophole in SBZL's argument, which means that their
result is actually incorrect.   As we show in
Appendix~\ref{ap:Zeilinger}, by means of a counter-example, 
\emph{operational} non-contextuality is \emph{not} inconsistent
with the empirical predictions of quantum mechanics.   The
modified argument  we give in Section~\ref{sec:ExtendKS2}
resolves this difficulty.  Lastly, the result we prove implies a
significantly different conclusion, as to the conditions which
must be satisfied in order refute non-contextual theories
experimentally.

Sections~\ref{sec:WhatObs}--\ref{sec:ExtendKS2} contain the main
part of our argument.  In
Sections~\ref{sec:POVM}--\ref{sec:EmpEquiv} we address a number of
related questions.

In Section~\ref{sec:POVM} we briefly discuss the POVM arguments
given by Kent~\cite{Kent1} and Clifton and Kent~\cite{Kent2}.

In Section~\ref{sec:Seq} we give an improved version of the
argument in Appleby~\cite{me3}.  In the first place we have
improved the argument so as to take account of the points made in
Section~\ref{sec:WhatObs} of this paper.  In the second place, the
version we give now does not involve joint measurements
of non-commuting observables.  In the third place, we have
strengthened the argument, using ideas derived from the subsequent
work of SBZL.

In Section~\ref{sec:EmpEquiv} we discuss the question, whether
there actually exists a \emph{complete} theory of MKC type
which is empirically
equivalent to quantum mechanics.  In particular, we discuss a
recent claim by Cabello~\cite{Cabel2}, that theories of this type
make experimentally testable predictions which conflict with
those of quantum mechanics.  We argue that, although Cabello makes
some very pertinent points, the question remains open.

\section{What is ``Actually'' Measured?}
\label{sec:WhatObs}
MKC's argument partly depends on a  misconception, regarding
non-ideal quantum measurements.   We begin our critical discussion
by clarifiying this point.  The criticisms which follow are
preliminary to our main critical argument, contained in the next
section.

MKC postulate that there are many observables which it is
physically impossible to measure.  
If an experimenter attempts to measure an observable $\hat{A}$
in this forbidden set, then  what is
\emph{actually} measured is a slightly different observable
$\hat{B}$.   The  measurement reveals
the pre-existing value of the observable $\hat{B}$ which is
\emph{actually} measured.

It can be seen from this that MKC tacitly assume:
\begin{list}{}
            {
             \setlength{\leftmargin}{1.174 in}
             \setlength{\rightmargin}{0.25 in}
             \setlength{\labelwidth}{1 in}
             }
\item[\textbf{Property 1}] For a given finite precision
measuring apparatus there is a single, uniquely defined
observable, which is the only observable that is ``actually''
measured.  
\item[\textbf{Property 2}] The  nominal
observable, which is recorded in the experimenter's notebook as
having been measured, is typically not the observable which is
``actually'' measured.
\end{list}
We will show that these assumptions are unjustified.

Before proceeding further, let us clarify the meaning of the term
``nominal observable'', as it appears in the above statement. 
The term essentially corresponds to Simon
\emph{et al}'s~\cite{Zeil1} ``switch position''.   However, unlike
Simon
\emph{et al} we are not appealing to the concept of an
``operational observable'' (\emph{i.e.}\ the concept that
different instruments define different observables---see
Appendix~\ref{ap:Zeilinger}).  We are simply recognizing the fact
that the complete record of a measurement (quantum or classical)
must include, not only a specification of the value obtained, but
also a specification of the observable measured.  Suppose, for
instance, that an experimenter makes 1,000 different measurements
of 1,000 different observables.  If the experimenter simply writes
down a list of 1,000 numbers, without any indication as to how
those numbers were obtained, then the record will obviously be
useless (this point, that the outcome of a measurement [quantum or
classical] is, not simply a \emph{value}, but a determinate
\emph{proposition}, will play a central role in the argument of
the next section).

It should  be stressed that, although we refer, for
convenience, to the actions of a human experimenter,  the same
point would apply if the measurements were performed by a
completely arbitrary ``information gathering and utilizing
system''~\cite{Gell}.  If the system is to utilize the
information it acquires then it must  record, in its  memory
area, a specification of the observables measured, as well as the
values obtained.

The nominal observable need not be specified precisely.  Rather
than  recording the information ``observable $\hat{S}_z$ was
measured and value $+1/2$ obtained'', one might instead record the
information  ``observable ${\mathbf n} \cdot \hat{\mathbf S}$ was
measured, for  ${\mathbf n} \in U$, and value $+1/2$ obtained''
(where $U$ is  some subset of the unit 2-sphere).  In the following
we will assume that the nominal observable is specified precisely,
because this is legitimate, and because it
happens to be the usual practice.    However,
the argument could easily be modified so as to allow for the
possibility that the nominal observable is only specified
partially.

Let us now consider MKC's assumption that, for each quantum
measurement, there is one, and only one observable which is
``actually'' measured.  
The fact that this assumption is unjustified becomes apparent when
one takes into view (as MKC do not) the detailed physical
implementation of an abstract quantum measurement scheme.

\begin{figure}[htb]
\epsffile{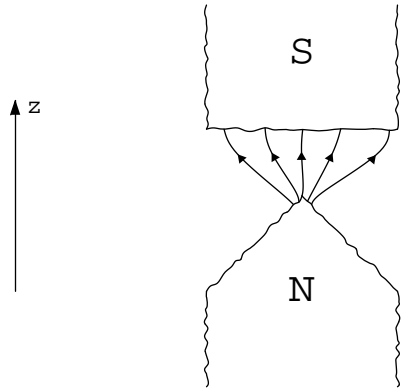}
\caption{Schematic illustration of the Stern-Gerlach arrangement. 
The surface irregularities of the pole pieces, and asymmetries in
the environment, mean that the field is not perfectly symmetric. 
There is no single direction  which represents the
alignment of the field more truly than any  other.   It 
follows that there is no single spin component which is measured
more truly than any other.}
\label{fig:SternGerlach}
\end{figure}

Consider, for example, a
non-ideal measurement of the spin component ${\mathbf n} \cdot
\hat{\mathbf S}$ using a Stern-Gerlach apparatus.  The vector
${\mathbf n}$ is determined by the axis of symmetry of the
arrangement.  MKC take it that, in such a case, although the axis
is not precisely known to the experimenter, there does actually
exist a single, sharply defined axis of symmetry; and that there
correspondingly  exists a single, sharply defined spin component
${\mathbf n} \cdot
\hat{\mathbf S}$ which is the \emph{only} spin component that is
actually measured.  However, this is to overlook the fact
that the magnetic field will not, in practice, be perfectly
symmetric (due  to irregularities in the shape of the
pole-pieces, asymmetries in the placement of surrounding objects,
\emph{etc.}).  As illustrated
in Fig.~\ref{fig:SternGerlach} the numerous slight departures from
perfect symmetry introduce some unavoidable ``fuzziness'' into the
concept ``axis of symmetry of the field''.   Under these
conditions there is no \emph{single} vector
${\mathbf n}$ which represents  the axis of symmetry more truly
 than any other.  This being so, there are no obvious grounds for
picking out any single spin component as the only component
which is ``actually'' measured.

A measurement~\cite{NotePred} of ${\mathbf n} \cdot \hat{\mathbf
S}$ is a process which discriminates the
eigenstates of ${\mathbf n} \cdot \hat{\mathbf
S}$.  Given an
unknown eigenstate of ${\mathbf n} \cdot \hat{\mathbf S}$, an
ideal measuring instrument will inform the experimenter, with
\emph{certainty}, which particular eigenstate it was.  A non-ideal
measuring instrument will inform the experimenter, with \emph{high
probability}, which particular eigenstate it was.  This is the
definition of a quantum measurement
process~\cite{NotePred}:  if a process performs the function, of
discriminating the eigenstates of an operator, then it is a
measurement of that operator, ideal or non-ideal as the case may
be.

Of course, it is seldom, if ever the case that a real laboratory
instrument is strictly ideal, and the Stern-Gerlach arrangement is
no exception to this  rule.  It is important, furthermore,
to note that, not only does a Stern-Gerlach arrangement fail to
perform an ideal measurement of the nominal spin component, which
is recorded in the experimenter's notebook as having been measured;
it fails to perform an ideal measurement of any other spin
component either (for a discussion of the unavoidable sources of
non-ideality in Stern-Gerlach measurements see Busch \emph{et
al}~\cite{BuschBk}).  The arrangement may, however, be used to
perform \emph{non-ideal} measurements (which is the most that can
reasonably be demanded of a real, laboratory instrument).

We can  characterize the degree of
non-ideality in quantitative terms by considering the probability
that the process will fail to correctly  identify a given
eigenstate.  Suppose that the arrangement is  used to measure
the component 
${\mathbf n} \cdot \hat{\mathbf S}$ for a spin-$1/2$ particle. 
 Let
$p_{+-}({\mathbf n})$ (respectively $p_{-+}({\mathbf n})$) be the
probability that, when the system particle is initially in the
spin up (respectively spin-down) eigenstate of
${\mathbf n} \cdot \hat{\mathbf S}$ the measurement outcome is
$-1/2$ (respectively $+1/2$).  Then it may be said that, the
smaller these probabilities are, the more nearly
ideal~\cite{NoteIdeal} the measurement.

There is no vector ${\mathbf n}$ for which 
$p_{+-}({\mathbf n})=p_{-+}({\mathbf n})=0$ (see Busch \emph{et
al}~\cite{BuschBk}).  There are, however, vectors ${\mathbf n}$
for which  the failure probabilities
$p_{+-}({\mathbf n})$,
$p_{-+}({\mathbf n})$ are both small.  If ${\mathbf n}$ satisfies
this condition the arrangement can be used to discriminate the
eigenstates of 
${\mathbf n} \cdot \hat{\mathbf S}$ with a high degree of
reliability.  In other words, it can be used to perform  non-ideal
measurements of
${\mathbf n}
\cdot
\hat{\mathbf S}$.

The failure probabilities 
$p_{+-}({\mathbf n})$,
$p_{-+}({\mathbf n})$ vary continuously with ${\mathbf n}$.  This
means that, if the probabilities are small for one vector
${\mathbf n}$, then they will also  be small for every other
vector which is close to ${\mathbf n}$.  It follows that, for a
given arrangement, there are infinitely many different spin
components which are \emph{all} non-ideally 
measured.  Furthermore the class of observables which are
non-ideally  measured includes the nominal observable (provided
it is correctly recorded, and provided the
instrument does not malfunction). 

We began our discussion of the Stern-Gerlach arrangement by
noting that, in practice, the arrangement will not have a sharply
defined axis of symmetry.  It can  now be seen  that, even if the
symmetry were exact, MKC's assumption, that there is only one
observable that is ``actually'' measured, would still not be
justified.  It is true that there might (perhaps) then be a
well-defined, natural  sense in which  the
measurement of
${\mathbf n} \cdot \hat{\mathbf S}$ was ``most nearly ideal'' when
${\mathbf n}$ was parallel to the axis of symmetry.  But it would
not follow that a neighbouring component
${\mathbf n}'
\cdot
\hat{\mathbf S}$, defined by a vector ${\mathbf n}'$ which was
 not exactly parallel to the axis of symmetry, was not 
``actually'' measured
\emph{at all}.

  A process does not have to be precisely optimal in
order to count as a measurement.  
If the probabilities  $p_{+-}({\mathbf n})$ and
$p_{-+}({\mathbf n})$ become slightly larger, then that simply
means that the process does not discriminate eigenstates quite so
efficiently as before.  It does not mean that the process thereby
ceases to \emph{be} a measurement.

 The real photon
detectors which have been constructed to date are  imperfect, in
that there is a non-zero probability that the detector will
fail to register the presence of a photon.  This does not mean
that a real photon detector does not ``actually'' detect 
photons at
all.   Furthermore, a photon detector whose performance
becomes  slightly degraded, so that it is no
longer quite so efficient as before, does not, on that account,
cease to \emph{be} a photon detector.

In general, however, there is no unambiguous, non-arbitrary,
physically well-motivated way to define the concept ``unique spin
component for which the measurement is most nearly ideal''.   For
instance, minimizing the function 
$(1/2)\bigl(p_{+-}({\mathbf n})+ p_{-+}({\mathbf n})\bigr)$, and
minimizing the function $\bigl((1/2)\left(p_{+-}({\mathbf
n})^2 + p_{-+}({\mathbf n})^2\right)\bigr)^{1/2}$ leads to
definitions  of this concept which are, in general, incompatible
(quite apart from the fact that the minima may not be unique). In
the general case there are \emph{many} spin components which all
have an equally valid claim to the status ``spin component for
which the measurement is most nearly ideal''.

Thus far we have been considering the  case of
the Stern-Gerlach apparatus.  However, it is shown in
Appendix~\ref{ap:NonIdeal} that the above criticisms
apply much more generally, to the case of any approximate von
Neumann measurement (\emph{i.e.}\ any non-ideal measurement
described by a unitary operator which is close to a unitary
operator describing a von Neumann measurement).  This provides
strong grounds for the assertion that Properties~1 and~2 assumed
by MKC  should be replaced by
\begin{list}{}
            {
             \setlength{\leftmargin}{1.174 in}
             \setlength{\rightmargin}{0.25 in}
             \setlength{\labelwidth}{1 in}
             }
\item[\textbf{Property $\bbox{1'}$}] 
For a given finite precision measuring apparatus  there are
infinitely many different observables which are non-ideally 
measured.
\item[\textbf{Property $\bbox{2'}$}]
The  nominal observable is among the observables which are
non-ideally measured
(provided the nominal observable is  recorded correctly, and
provided there is no malfunction in the measuring apparatus).
\end{list}

It may  be worth remarking that these statements 
are also valid classically.   Consider, for example a classical
measurement of the velocity component
$v_z$ of some macroscopic object, using the Doppler shift of the
radiation it emits.   In order to perform this measurement it is
necessary that the line running from source to detector should be
parallel to the
$z$-axis.  The fact that source and detector are both extended
objects, which are not perfectly symmetric, means that, in
practice, this line is not even precisely 
\emph{defined}, let alone precisely \emph{known}.  
However, this does not imply that 
$v_z$ may not be the component which is ``actually''  measured. 
Rather, it is to be regarded as one of the sources of error
in the  measurement of $v_z$.   A classical measurement
of a classical observable $A$ is a process which reveals, with a
high degree of reliability, and to a high degree of accuracy, the
pre-existing value of $A$.   On this definition the procedure
described effects a non-ideal classical measurement of ${\mathbf n}
\cdot {\mathbf v}$ for every ${\mathbf n}$ which is
nearly (but perhaps not precisely) parallel to the $z$-axis.

The effect of the argument in this section is to restore the
natural assumption of most physicists, that the observable which an
experimenter records as having been non-ideally measured is
 also measured in fact (provided the experimenter does
not make a mistake, and provided there is no malfunction in the
measuring apparatus).

Let us now consider the bearing which this has on MKC's argument.
Our criticisms do not invalidate MKC's
postulate, that a measurement of $\hat{A}$  may reveal the
value of some other,  neighbouring observable
$\hat{B}$.   They do, however, show that MKC cannot justifiably
claim that $\hat{B}$ is the only observable which is ``actually''
measured, and that $\hat{A}$ is not really measured at all.   MKC
cannot even claim that $\hat{B}$ is the uniquely defined 
observable for which the measurement is most nearly ideal.  In
general, the observable $\hat{B}$ has \emph{no} preferred
status.  Properly understood, MKC's postulate comes to this:  that
the hidden dynamics somehow selects an essentially
\emph{arbitrary} observable
$\hat{B}\approx\hat{A}$, whose value it then reveals.

The point is important because, once MKC's proposal is formulated
correctly, it becomes  easier to see the fallacy in
their argument.

\section{Nullification of the Nullification}
\label{sec:ExtendKS1} 
This section contains the main part of our critical argument.

The KS theorem shows
\begin{list}{}
            {
             \setlength{\leftmargin}{1.174 in}
             \setlength{\rightmargin}{0.25 in}
             \setlength{\labelwidth}{1 in}
             }
\item[\textbf{Proposition 1}] No hidden variables interpretation
can have the property that,  for every observable
$\hat{A}$,  an \emph{ideal} measurement of 
$\hat{A}$  always  reveals the pre-existing value of
$\hat{A}$. 
\end{list}
MKC's claim, when appropriately re-formulated, so as to remove any
unjustified appeal to the concept of ``the unique observable
which is actually measured'', is that this proposition is
``nullified'' by
\begin{list}{}
            {
             \setlength{\leftmargin}{1.174 in}
             \setlength{\rightmargin}{0.25 in}
             \setlength{\labelwidth}{1 in}
             }
\item[\textbf{Proposition 2}]  There do exist hidden
variables interpretations  with the property that,  for
every  observable
$\hat{A}$,  a \emph{finite precision}  
measurement of $\hat{A}$ always reveals the pre-existing
value of some unknown observable $\hat{B}$ in a small
neighbourhood of $\hat{A}$.   
\end{list}
In considering this claim the first problem we face is
that the term 
``nullified'' is not part of the standard lexicon of science and
mathematics, and so its meaning is potentially ambiguous. 
Propositions~1 and~2 are logically consistent, so there can be no
question of the KS theorem actually being refuted.   It is not
immediately apparent what, exactly, is meant by the claim
that Proposition~2 ``nullifies'' Proposition~1, without
contradicting it. 

In this connection it should be noted that the reference to
finite precision measurements, on which MKC
themselves place  much emphasis, is actually irrelevant.  MKC
postulate that, for each non-ideal measuring apparatus, the hidden
dynamics somehow arbitrarily selects an observable 
$\hat{B}$, close to the nominal observable, whose value the
measurement  then reveals.   One could, with equal justification,
postulate that the same is true for each \emph{ideal} measuring
apparatus: implying
\begin{list}{}
            {
             \setlength{\leftmargin}{1.174 in}
             \setlength{\rightmargin}{0.25 in}
             \setlength{\labelwidth}{1 in}
             }
\item[\textbf{Proposition $\bbox{2'}$}]  There do exist hidden
variables interpretations  with the property that,  for
every  observable
$\hat{A}$,  an \emph{ideal}  
measurement of $\hat{A}$ always reveals the pre-existing
value of some unknown observable $\hat{B}$ in a small
neighbourhood of $\hat{A}$.   
\end{list}
If  the KS theorem is nullified by Proposition~2, then it must,
presumably, also be nullifed by Proposition~$2'$.   It would seem
to follow that such nullification as there might be cannot
be attributed specifically to the finite precision. 

In the following we will take the claim to be that the MKC models
invalidate what had previously been regarded as the key physical
implication of the KS theorem:  namely, the implication that
quantum measurements cannot systematically be interpreted as
measurements in the classical sense. In the MKC models a
measurement does not always reveal the value of the nominal
observable.  It does, however, always reveal the value of an
observable which is extremely close to the nominal observable.  
MKC assume that is enough to recover the classical concept of
measurement.

The question we now have to decide is whether MKC are correct to
assume that a process which does not reveal the pre-existing value
of the nominal observable may still count as a 
measurement in the classical sense.  We will argue that they are
not.

The crucial point to realize is that a classical measurement is
not \emph{simply} a process which reveals a pre-existing value. 
 Rather, it is a process which
reveals a pre-existing piece of classical information, represented
by a proposition  of the form ``observable
$A$ took value $x$''.  For this to be true  it is essential that
the observable $A$ and the value $x$ both be adequately specified. 
A process which reveals the value of some 
\emph{completely unknown} observable does not reveal any classical
information, and is not a classical measurement. 

The following example may serve to illustrate this point. 
Suppose that an experimenter is given a sealed box, containing
100 individually labelled rods.  Suppose that, for each integer
$n=1,\dots,100$ the box contains exactly one rod of length
$n\ {\rm cm}$.   Suppose that the experimenter, without looking
inside the box, uses a random number generator to select a
particular integer $1 \le n_0 \le 100$.  It could be said that
this procedure reveals the length of one of the rods in the box. 
However, it does not reveal the identity of that rod. 
Consequently, the process does not reveal any classical
information, and so it cannot be considered a classical
measurement.

It may, at first sight, seem that these remarks do not apply to
the situation envisaged by MKC since, in their models, the
observable
$\hat{B}$, whose value is revealed by the measurement, is not
completely unknown.  In fact, it is assumed that $\hat{B}$ is
extremely close to the nominal observable $\hat{A}$:  so there is
a sense in which $\hat{B}$ is specified very precisely.   However,
it is not specified precisely \emph{enough}.

The point here is that the MKC valuations are radically
discontinuous.  If the valuation could be assumed  continuous
in the vicinity of $\hat{A}$, then
a process which revealed the value of some neighbouring
observable $\hat{B}$ could be regarded as a measurement in the
classical sense.   However, there is in fact always a region on
which the valuation is  highly discontinuous, and so this
assumption would not generally be justified. 

Consider, for example, the colourings of the unit 2-sphere $S^2$
discussed by Meyer~\cite{Meyer} (and described in more detail by
Havlicek \emph{et al}~\cite{Havli}).   These colourings have the
property  that,
given any ${\mathbf n} \in S^2$, every neighbourhood of ${\mathbf
n}$ contains infinitely many points evaluating to 0, and
infinitely many points evaluating to 1.  Suppose, now, that one
performs a finite precision measurement of the nominal observable
$({\mathbf n} \cdot \hat{\mathbf S})^2$ and obtains (say) the
value 1.  Then one knows that, there is a point ${\mathbf n}'$
in some small  neighbourhood $U$ of ${\mathbf n}$ which evaluates
to 1.  However, $U$ contains infinitely many points evaluating
to 1, and infinitely many points evaluating to 0.  The process
does not reveal any more information, as to which particular point
takes which particular value, than could be obtained by tossing a
coin.  Consequently, it cannot be considered a classical
measurement.  It is not a classical measurement for  the same
reason that random number generators cannot be used to make
classical measurements of length.    

Essentially the same criticism applies to an arbitrary model of MKC
type.  In the general case the  highly discontinuous behaviour
just discussed may not occur everywhere, on the whole of $S^2$. 
However, there will always be a non-empty open subset of $S^2$ on
which such behaviour occurs.  

The detailed proof of this statement is given in
Appendix~\ref{ap:Colour}.  We will confine ourselves here to
summarising the main points.

Consider an arbitrary MKC colouring $f\colon S^2_0\to \{0,1\}$. 
Here 
$S_0^2$ is any dense, KS-colourable subset of the unit 2-sphere
$S^2$ having the  property that the set of triads
contained in 
$S_0^2$ is dense in the space of all triads (MKC 
assume that $S_0^2$ is also countable; however  the argument which
follows does not depend on this assumption).  

 We
define the discontinuity region
$D\subseteq S^2$ to consist of those points 
${\mathbf n} \in S^2$ with the
property that each neighbourhood of ${\mathbf n}$ contains
infinitely many points evaluating to 0, and infinitely many points
evaluating to 1 (it is \emph{not} assumed that 
${\mathbf n}$ itself $\in S_0^2$).  We
define the continuity region
$C\subseteq S^2$ to consist of those points ${\mathbf n} \in
S^2$ with the property that $f$ is continuous on $U\cap S^2_0$,
for some neighbourhood $U$ of ${\mathbf n}$.  In  the case of the
Meyer colourings $C=\emptyset$ and $D=S^2$.  However, this is
not true generally.   

It is readily verified that these regions partition $S^2$ (the
\emph{complete} unit 2-sphere, not just  $S^2_0$) into two
disjoint subsets: 
$C\cup D = S^2$ and
$C
\cap D =
\emptyset $.   It is also readily verified that $C$ is
open and
$D$ is closed.

The key result, proved in Appendix~\ref{ap:Colour}, is that
$C$ is not only open, but also KS-colourable.  Furthermore, the
KS-colouring
$f$ defined on
$C \cap S^2_0$ uniquely extends to a continuous, induced
KS-colouring
$\bar{f}\colon C \to \{0,1\}$.

We show in Appendix~\ref{ap:Colour} that this result places 
significant constraints on the minimum size of the discontinuity
region. In the first place,  $D$
must have non-empty interior (\emph{i.e.}\ it must contain a
non-empty open subset---implying that it cannot simply consist of
a set of isolated points or lines).  In the second place,  there is
a fixed, model-independent, non-zero lower bound on the solid
angle subtended by $D$.

Let us now consider  the bearing of these results on MKC's claim,
to have ``nullified'' the KS-theorem. 
The argument we gave above (in connection with the Meyer
colourings) shows that the $f$-value assigned to a specific point
$\in D
\cap S^2_0$ cannot be ascertained  by any 
quantum measurement and is, in this sense, inaccessible.  On the
other hand the $f$-value assigned to a specific point $\in C
\cap S^2_0$ can be ascertained by a
quantum measurement,
and so it is accessible (provided the observable whose
value is revealed is sufficiently close to the nominal observable).
  It is easy to see that the same is also true
of the $\bar{f}$-value assigned to a specific point $\in C$ (where
$\bar{f}$ is the induced KS-colouring defined above).

From an empirical point of view it is not the $f$-values defined 
on $S_0$ which are  significant, but the
$\bar{f}$-values defined on $C$. It is only when ${\mathbf n} \in
C$ that a quantum measurement of
$({\mathbf n} \cdot \hat{\mathbf S})^2$ can be interpreted as a
measurement in the classical sense, which reveals the
corresponding $\bar{f}$-value (provided the observable whose
value is revealed is sufficiently close to the nominal observable).

The fact that there exist non-empty, KS-colourable open
subsets of
$S^2$ is trivial (consider, for example, the
cone
$C=\{{\mathbf n}
\in S^2\colon |n_z| > 1/\sqrt{2}\}$).  There is no  sense in
which this fact can validly be said to nullify the KS theorem.

In conclusion, it can be seen that the finely calculated approach
of MKC, where one marks in individual points using a pencil,
gets no nearer to ``nullifying'' the KS theorem
than does a  cruder approach, where one paints in
whole open regions using a brush.  It can also be seen that a
would-be nullifier gets no nearer to his/her goal by
postulating that a measurement  reveals the value of an
observable which is typically not the nominal observable.

\section{Nullifying Made Easy}
\label{sec:NullFurther}
This section contains some supplementary considerations.  Its
purpose is  to reinforce the point made in the last section that,
for a model to nullify the KS theorem, rather more is required
than the bare fact, that a measurement always reveals the
pre-existing value of
\emph{something}.

If the KS theorem could be nullified in the manner that MKC
propose, then one could achieve that end more
straightforwardly, without appealing to MKC's sophisticated
set-theoretic constructions, by means of one of the following, much
less sophisticated models.

Since it is a hidden variables theory that is in question, Ockham's
razor does not apply, and so we are free to multiply hypothetical
entities at will.  Let us accordingly postulate that, corresponding
to each observable $({\mathbf n} \cdot \hat{\mathbf
S})^2$ of the conventional theory, there exist two beables $s_{A}
({\mathbf n})$ and $s_{B}
({\mathbf n})$.  Each  beable is assigned the value 0 or 1.   We
postulate that these values fluctuate randomly, and independently,
but in such a way that, for each ${\mathbf n}$, at every time,
exactly one of the beables  $s_{A}
({\mathbf n})$, $s_{B}
({\mathbf n})$ is assigned the value 0, and exactly one is
assigned the value 1.  In every other respect the model coincides
with conventional quantum mechanics.  We are thus assuming that the
values taken by the beables have no causal influence on the
outcome of a measurement---or, for that matter, on anything else. 
Their function is simply to exist.  We will refer to this as
model~1.

It is trivially true that, in model~1, the effect of measuring 
$({\mathbf n} \cdot \hat{\mathbf
S})^2$ is to reveal the pre-existing value of one of the
two quantities  
$s_{A} ({\mathbf n})$ or $s_{B}
({\mathbf n})$.  This is to be compared with the situation in the
MKC models, where the effect is to reveal the value of one out of
an infinite set of vectors contained in some small neighbourhood
of ${\mathbf n}$.  It would consequently seem that, if one is to
accept that the MKC models nullify the KS theorem, then one must
accept that model~1 nullifies it equally well.

Of course, it is intuitively evident that the KS theorem
is not really  nullified by model~1. 
However, it is worth examining the reason for this. 

Model~1 does not nullify the KS theorem because, even though the
measurement reveals a pre-existing value, the experimenter does
not learn anything thereby.   The experimenter already knows,
before the measurement, just from the way  the model is
defined, that exactly one of the two quantities 
 $s_{A} ({\mathbf n})$ or $s_{B}
({\mathbf n})$ takes the value 0.  The experimenter does not know
more after the measurement.  The procedure cannot be used to
acquire additional knowledge concerning  
$s_{A} ({\mathbf n})$, $s_{B}
({\mathbf n})$ and so it is not, in the classical sense, a
measurement of these quantities.  In short, model~1 does not
nullify the KS theorem for essentially the same reason that the MKC
models do not nullify it.

It is also illuminating to consider the following, slightly
different model, which we will refer to as model~2.  Model~2 is
the same as model~1, except that the quantities 
$s_{A} ({\mathbf n})$, $s_{B}
({\mathbf n})$ are not assumed to fluctuate.  Instead, it is
postulated that, for each ${\mathbf n}$, $s_{A} ({\mathbf n})$
always takes the value 0, and $s_{B}
({\mathbf n})$ always takes the value 1.

In model~2 a measurement of
$({\mathbf n}
\cdot
\hat{\mathbf S})^2$ not only reveals the pre-existing value of 
$s_{A} ({\mathbf n})$ or $s_{B}
({\mathbf n})$:  the experimenter even knows which of the two
quantities took this value.  Nevertheless, there is still no
nullification.
Suppose that an experimenter measures $({\mathbf n}
\cdot
\hat{\mathbf S})^2$ and obtains the value 0.  Then, on the
assumptions of model~2, the experimenter knows that $s_{A}({\mathbf
n})$ took the value 0.  However, the experimenter knew this
already, before carrying out the measurement.  The procedure has
not conveyed any
\emph{additional} information, regarding $s_{A}({\mathbf
n})$, and so it is not, in the classical sense, a measurement of 
$s_{A} ({\mathbf n})$.

The triviality of the constructions makes it evident that models~1
and~2 do not really nullify the KS theorem.  However, the
underlying fallacy is essentially the same as the fallacy
underlying the much more sophisticated arguments of MKC.   The
only difference is that, in the case of the MKC models, partly
because the constructions are so  much more sophisticated, the
fallacy is  less obvious.

\section{KS Theorem for Non-Ideal Measurements}
\label{sec:ExtendKS2}
In Sections~\ref{sec:WhatObs}--\ref{sec:NullFurther} we established
the negative result: that finite precision does not
``nullify'' the KS theorem.  The purpose of this section is to
establish the positive result:  that the KS theorem has a
natural generalization, applying to the non-ideal measurements
which might actually be performed, in the laboratory.

The original version of the KS theorem shows
\begin{list}{}
            {
             \setlength{\leftmargin}{1.174 in}
             \setlength{\rightmargin}{0.25 in}
             \setlength{\labelwidth}{1 in}
             }
\item[\textbf{Proposition 1}] No hidden variables interpretation
can have the property that,  for every observable
$\hat{A}$,  an \emph{ideal} measurement of 
$\hat{A}$  \emph{always}  reveals the  pre-existing
value of
$\hat{A}$. 
\end{list}
---meaning that ideal quantum measurements cannot systematically
be interpreted as ideal classical measurements, of the same
observables.

The generalized version of the KS theorem 
shows
\begin{list}{}
            {
             \setlength{\leftmargin}{1.174 in}
             \setlength{\rightmargin}{0.25 in}
             \setlength{\labelwidth}{1 in}
             }
\item[\textbf{Proposition $\bbox{1'}$}] No hidden variables
interpretation can have the property that,  for every observable
$\hat{A}$,  a 
\emph{near-ideal} measurement of 
$\hat{A}$ always has a \emph{high probability}  of revealing a 
\emph{good
approximation} to the pre-existing value of
$\hat{A}$. 
\end{list}
---meaning that
near-ideal quantum measurements cannot systematically be
interpreted as near-ideal classical  measurements, of the same
observables.

Proposition~$1'$ is a qualitative statement, since the terms
  ``near-ideal'', ``high
probability'' and ``good approximation'' are not sharply
defined.  In the following we will prove an
inequality which gives unambiguous, quantitative expression to
Proposition~$1'$, as it applies to the standard example of a
spin~1 system (see Inequality~(\ref{eq:NonIdealKSResult1}) below).

Proposition~$1'$ does not exclude the possibility that a
near-ideal quantum measurement may reveal the pre-existing value
of something other than the nominal observable.   As we saw in
Sections~\ref{sec:ExtendKS1} and~\ref{sec:NullFurther} this
possibility has no bearing on the essential point of the KS
theorem, which is to show that there is a radical difference
between the quantum mechanical and classical concepts of
measurement.

Proposition~$1'$ does not \emph{only} exclude the possibility of
interpreting quantum measurements as \emph{ideal} classical
measurements.  This is important, since there would otherwise be
no conflict with classical assumptions.  One does not classically
expect a real, laboratory measurement to be strictly ideal (in the
classical sense).

The argument which follows was  inspired by the work of
Simon~\emph{et al}~\cite{Zeil1,Simon2} and Larsson~\cite{Lars}
(SBZL).    However, we do not use their ``operational'' definition
of contextuality.  This makes the relationship with the original KS
theorem much clearer.   Furthermore, there is a loophole in SBZL's
argument, which means that the result stated by them is actually
invalid (\emph{operational} non-contextuality is \emph{not}
inconsistent with the empirical predictions of quantum mechanics).
Lastly, the result we prove implies  a significantly different
conclusion, as to the conditions which must be satisfied in order
to refute non-contextual theories experimentally.  For further
discussion of SBZL's approach see the end of this section, and
Appendix~\ref{ap:Zeilinger}.

Consider the standard case of a spin-1 system, with spin vector
$\hat{\mathbf S}$.  Consider an apparatus which can be used to
perform a joint, possibly non-ideal measurement of the 
observables 
$({\mathbf e}_1 \cdot \hat{\mathbf S})^2$,  
$({\mathbf e}_2 \cdot \hat{\mathbf S})^2$,  
$({\mathbf e}_3 \cdot \hat{\mathbf S})^2$ for any  
orthonormal triad $\tau=\{{\mathbf e}_1, {\mathbf e}_2, {\mathbf
e}_3\}$.

We are here assuming that it is  physically possible to
(non-ideally) measure
\emph{every} 
 triad $\tau$---contrary to what is
postulated by MKC.  The justification for this assumption is that,
if it is physically possible to (non-ideally) measure \emph{one}
triad, then it follows from rotational invariance, together with
the argument in Section~\ref{sec:WhatObs} and
Appendix~\ref{ap:NonIdeal} of this paper, that it is physically
possible  to (non-ideally) measure every other triad.  Suppose,
for instance, that one is given an apparatus which (non-ideally)
measures the triad $\tau$, and one wants to measure the triad
$\tau'$ which is obtained by rotating $\tau$ through the Euler
angles $\theta, \phi, \psi$.  Then one simply rotates the
apparatus through these angles.   The fact that the rotation
cannot be performed with infinite precision does not mean that the
apparatus may not measure $\tau'$ at all. It only means that the
measurement of $\tau'$ may be less nearly ideal than would
otherwise be the case (\emph{i.e.}\ it may discriminate the
eigenstates of $\tau'$ less reliably than would otherwise be the
case). For further discussion of the practical realization of such
measurements see Swift and Wright~\cite{Swift}.

Let $p_{\rm q}(\tau,\psi)$ be the probability that a measurement
of $\tau$ results in one of the ``illegal'' combinations
$000, 001,010, 100, 111$, given that
the  system was prepared in the state $\psi$. Define
\begin{equation}
\epsilon_{\rm q} = \sup_{\tau, \psi} \bigl( p_{\rm
q}(\tau,\psi)
\bigr)
\label{eq:EpsQDef}
\end{equation}
The quantity $\epsilon_{\rm q}$  partially characterizes the
degree of non-ideality of the quantum measuring device (the
characterization is only partial because the condition
$\epsilon_{\rm q}=0$, though necessary, is not sufficient for
strict ideality).  It  corresponds to SBZL's quantity $\epsilon$
(\emph{mutatis mutandi}).

Let $f_{\lambda}
({\mathbf n})$ be the pre-existing value of  
$({\mathbf n} \cdot \hat{\mathbf S})^2$ when the hidden state is
$\lambda$ (it should be stressed that $f_{\lambda}$ is defined for
every
${\mathbf n}$, and so it is \emph{not} a KS-colouring).  Let 
$p_{\rm c} (\tau, \lambda)$ be the probability that, with the
system initially in the hidden state $\lambda$,  a measurement of
$\tau$ fails to reveal the $f_{\lambda}$-values of $\tau$. Define
\begin{equation}
\epsilon_{\rm c} = \sup_{\tau, \lambda} \bigl( p_{\rm
c} (\tau,\lambda)
\bigr)
\label{eq:EpsCDef}
\end{equation}
The quantity $\epsilon_{\mathrm c}$ characterizes the degree of
non-ideality of the apparatus regarded as a classical measuring
device, whose function is to reveal pre-existing values. It should
be noted that $\epsilon_{\mathrm c}$ depends on the particular
hidden variables theory considered.  Unlike $\epsilon_{\rm q}$ it
has no empirical significance.

We will now prove 
\begin{equation}
 N \epsilon_{\rm q} + \epsilon_{\rm c} \ge 1
\label{eq:NonIdealKSResult1}
\end{equation}
where $N$ is the smallest number of triads  which do not admit a
consistent colouring.   

This inequality shows that $\epsilon_{\rm q}$ and $\epsilon_{\rm
c}$ cannot both be made arbitrarily small.  It thus gives precise,
quantitative expression to the qualitative statement of
Proposition~$1'$.

Proposition~$1'$ is a general statement, so it allows for the
possibility that a measurement may reveal a good approximation to
the pre-existing value, without revealing it exactly.  In the
case considered here there are only two possible values, and so
this contingency cannot arise.

To prove this inequality, consider the probability $p_{\rm i}
(\tau, \lambda)$ that, when the hidden state is $\lambda$, a
measurement of $\tau$ gives one of the ``illegal'' results
$000, 001,010, 100, 111$.  We have, for all $\psi$,
\begin{equation}
p_{\rm q}(\tau, \psi) = \int_{\Lambda} p_{\rm i}(\tau,
\lambda) \, d \nu_{\psi}
\label{eq:PqTermsPi}
\end{equation}
where $\Lambda$ is the hidden state space (of the system by
itself), and
$\nu_{\psi}$ is the  probability measure on $\Lambda$ which 
corresponds to the quantum state
$\psi$.

Now choose a KS-uncolourable set which contains the smallest
possible number of distinct orthornormal triads.  Let $\tau_1, 
\dots, \tau_N$ be an enumeration of these triads.  By
construction there exists, for each  $\lambda$, an
index $r_{\lambda}$ such that the $f_{\lambda}$-values of
$\tau_{r_\lambda}$ are ``illegal''.  It is then a straightforward
consequence of the definitions that
\begin{equation}
1-p_{\rm i}(\tau_{r_{\lambda}}, \lambda) \le
p_{\rm c} (\tau_{r_{\lambda}}, \lambda)
\end{equation}
implying
\begin{equation}
p_{\rm i}(\tau_{r_{\lambda}}, \lambda)\ge 1-\epsilon_{\rm c}
\end{equation}
for all $\lambda$. Taking this result in conjunction
with Eqs.~(\ref{eq:EpsQDef}) and~(\ref{eq:PqTermsPi}) we deduce
\begin{equation}
N \epsilon_{\rm q}
\ge \sum_{r=1}^{N} p_{\rm q}(\tau_r , \psi)
= \int \biggl( \sum_{r=1}^{N} p_{\rm i} (\tau_r,\lambda) \biggr)\,
d\nu_{\lambda}^{\psi}
\ge \int p_{\rm i} (\tau_{r_{\lambda}}, \lambda)\,
d\nu_{\lambda}^{\psi}
\ge 1-\epsilon_{\rm c}
\end{equation}
---which is the result claimed.

It is worth remarking that this argument actually establishes
the stronger, state-dependent inequality
\begin{equation}
 N \epsilon_{\rm q} (\psi) + \epsilon_{\rm c} \ge 1
\end{equation}
where 
$
\epsilon_{\rm q} (\psi)=
\sup_{\tau} \bigl( p_{\rm q}(\tau,\psi)
\bigr)
$.

If the measurement \emph{faithfully} reveals pre-existing
values then Inequality~(\ref{eq:NonIdealKSResult1}) implies
$N \epsilon_{\rm q} \ge 1 $.   This is similar to the result
claimed by SBZL~\cite{Zeil1,Lars,Simon2}, that  $N \epsilon_{\rm
q} \ge 1$ in any theory which is operationally
non-contextual (substituting our $\epsilon_{\rm q}$ for their
$\epsilon$).  However, it is not the same because a theory which
does not faithfully reveal pre-existing values may still be
non-contextual in the sense of SBZL's operational definition.  
SBZL's  claim is actually incorrect, since there exist
operationally non-contextual theories with the property
that
$N
\epsilon_{\rm q} =0$ (see Appendix~\ref{ap:Zeilinger}).

Simon \emph{et al}'s~\cite{Zeil1} statement, that there is  a
conflict with classical assumptions as soon as $\epsilon_{\rm q} <
1/N$, also requires modification.  Suppose that $\epsilon_{\rm q}$
is only  slightly less than $<1/N$.  Then it is true that the
classical failure probability cannot be strictly zero.  However,
it may still be very small---which would be entirely consistent
with classical assumptions regarding a \emph{real} laboratory
instrument.  Indeed, the possibility is not excluded that
$0<
\epsilon_{\rm c}
\ll
\epsilon_{\rm q} < 1/N$---implying that the apparatus actually
functions much
\emph{better} when regarded from the classical point of view, as
an instrument for revealing pre-existing values, than it does when
regarded from the quantum point of view, as an instrument for
discriminating eigenstates.  It follows that, in order to
demonstrate an experimental conflict with the classical concept of
measurement, it is not enough simply to construct an apparatus for
which 
$\epsilon_{\rm q} < 1/N$.  Instead, one must reduce
$\epsilon_{\rm q}$ to the point where the lower bound on
$\epsilon_{\rm c}$ becomes $\sim 1$.

The purpose of this section was to establish the basic point of
principle, that the KS theorem does have a valid
generalization applying to non-ideal measurements.  For this
purpose it was sufficient to confine ourselves to the single
example of a spin~1 system.  However, it would clearly be
interesting to examine other examples.  Indeed, we have not shown
that Inequality~(\ref{eq:NonIdealKSResult1}) is the only, or even
the most useful inequality applying  a spin~1
system.   One would expect there to be many different ways to 
give  quantitative expression to the qualitative principle
represented by Proposition~$1'$.  This is a question which
requires further investigation.

\section{Kent and Clifton's POVM Argument}
\label{sec:POVM}
In this section we briefly consider the argument of
Kent~\cite{Kent1} and Clifton and Kent~\cite{Kent2}  to (as
they put it) ``rule out falsifications of non-contextual models
based on generalized observables represented by POV measures''. 
 We include this section for the sake of completeness.
A reader who is so inclined
may skip to the next section without loss of continuity.

A POVM (or positive operator valued
measure~\cite{BuschBk,Dav,Hol,Niels})  is an indexed set   of
positive operators
$\hat{F}_i$ with the property
$\sum_{i}
\hat{F}_{i}=1$ (following Kent and Clifton we take the set to be
finite).  Kent and Clifton take the view that to each distinct POVM
there corresponds a distinct generalized observable.  A generalized
observable, in effect, simply is a POVM.

Kent and
Clifton consider that a non-contextual theory of generalized
observables would be one in which generalized  measurements 
reveal the pre-existing truth values of the corresponding POVMs. 
Their argument to show that such theories exist is similar to the
argument which, they claim, nullifies the ordinary KS theorem. 
The objection to it is also similar.

Let ${\mathcal A}$ be the set of all positive operators.  Kent
and Clifton postulate that the POVM which is \emph{actually}
measured is always contained in a certain dense subset  ${\mathcal
A}_d
\subset {\mathcal A}$, which is colourable in a generalized
sense.   If
 $\hat{F}_i$ is the POVM that is  \emph{actually} measured, then
the outcome will be the unique index $\emph{i}_0$ for which 
$\hat{F}_{i_0}$ evaluates to ``true''.

The objection to this argument is essentially  the point we have
already  made, in Sections~\ref{sec:ExtendKS1} 
and~\ref{sec:NullFurther}.  The problem is that the
experimenter does not know the identity of the positive
operator whose truth value is revealed.  Suppose that an
experimenter measures the nominal POVM $\hat{F}_i$, and obtains
the result $i_0$.  Then s/he knows that there is another,
typically different POVM  
$\hat{F}'_i$ for which $\hat{F}'_{i_0}$ is ``true''.   The
experimenter knows that \emph{something} is ``true''.   However,
the experimenter does not know what that something is. 
Consequently, the procedure  is not genuinely informative.  If one
is told ``proposition $p$ is true'', but is not given any
indication what the symbol $p$ denotes, then one has not really
been told anything.     

Kent and Clifton consider  that
their POVM argument includes, as a special case, an alternative
proof  that non-ideal measurements of conventional observables can
be explained non-contextually~\cite{NotePOVM}.  Let us 
consider how this might work.

POVMs play an important  
role in
the mathematical description of  non-ideal measurements of
conventional  observables. 
Consider, for example, the non-ideal measurement of the
conventional observable $\hat{A} = \sum_{a} a |a\rangle \langle
a|$ described by Eq.~(\ref{eq:ProxVonNeu}) in
Appendix~\ref{ap:NonIdeal}.  Define
\begin{equation}
\hat{E}_a = |a\rangle \langle a|
+\sum_{b} \epsilon^{\vphantom{*}}_{b, a a} |a\rangle \langle b|
+\sum_{b} \epsilon^{*}_{b,aa } |b\rangle \langle a|
+\sum_{b_1, b_2 , c} \epsilon^{*}_{b_1,c a }
\epsilon^{\vphantom{*}}_{b_2, c a} |b_1\rangle \langle b_2|
\label{eq:POVMNonIdeal}
\end{equation}
It is readily verified that  the operators 
$\hat{E}_a$  constitute a POVM.  Their physical significance is
that, if the system is initially in the state $|\psi\rangle$, then 
$\langle \psi | \hat{E}_a | \psi \rangle$ is the probability
that the result of the measurement will be 
$a$.

Let us note, parenthetically, that it may be questioned whether the
concept of a generalized observable is
 appropriate in the case of POVMs which arise in this way,
as part of the mathematical description of  non-ideal
measurements of conventional observables~\cite{me3,Uffink}. 

Clifton and Kent consider that, in the case of the
measurement process described by Eq.~(\ref{eq:ProxVonNeu}), what
is \emph{actually} measured is, not really $\hat{A}$ at all, but
rather  the POVM
$\hat{E}_a$.  Of course, the unitary operator $\hat{U}$ in
Eq.~(\ref{eq:ProxVonNeu}), and the POVM 
$\hat{E}_a$ defined in terms of it,
could not, in practice, be known, with infinite precision. 
Consequently,
Clifton and Kent's POVM argument, when specialized to non-ideal
measurements of conventional observables, establishes the
following:  there exist models in which the effect of non-ideally
measuring an observable
$\hat{A}=\sum_{a} a |a\rangle \langle a|$ is to reveal the
pre-existing truth values of some unknown POVM $\hat{E}_a \approx
|a\rangle \langle a|$. Conceived as a way  of nullifying the KS
theorem this result is no more effectual than Proposition~2,
which we criticized in Sections~\ref{sec:ExtendKS1}
and~\ref{sec:NullFurther}.

\section{Contextuality of Sequential Analyzers}
\label{sec:Seq}
In this section we discuss the predictions of the MKC models
regarding sequential measurements, using three separate analyzers. 
If the arrangement is near-ideal, then 
the observable whose value is revealed by one analyzer  must depend
on the settings of the other two analyzers.  In other words, it
must depend on the overall experimental context.

This result serves to reinforce the conclusion reached earlier,
that the MKC models are highly non-classical.  It also raises the
question,  whether the MKC models are  capable
of reproducing the empirical predictions of quantum mechanics---as
 discussed in the next section.

The argument which follows is  an improved version of an
argument given in Appleby~\cite{me3}.  In the first place we have
improved the argument so as to take account of the points made in
Section~\ref{sec:WhatObs} of this paper.  In the second place, the
version we give now does not involve joint measurements
of non-commuting observables.  In the third place, we have
strengthened the argument, using ideas derived from the subsequent
work of SBZL.

\begin{figure}[th]
\epsffile{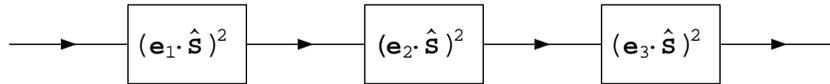}
\caption{Sequential measurement of
$({\mathbf e}_1 \cdot
\hat{\mathbf S})^2$, $({\mathbf e}_2 \cdot \hat{\mathbf
S})^2$, $({\mathbf e}_3 \cdot \hat{\mathbf
S})^2$.  The system passes through a succession of analyzers,
which measure each observable in turn.}
\label{fig:sequential}
\end{figure}

Consider sequential measurements of the observables   
$({\mathbf e}_1 \cdot \hat{\mathbf S})^2$,  
$({\mathbf e}_2 \cdot \hat{\mathbf S})^2$,  
$({\mathbf e}_3 \cdot \hat{\mathbf S})^2$, using the apparatus
illustrated in 
Fig.~\ref{fig:sequential}.  The apparatus consists of three
identical, non-ideal analyzers chained together.
$\hat{\mathbf S}$ denotes the spin vector of a spin 1 system, and
$\{{\mathbf e}_1, {\mathbf e}_2, {\mathbf e}_3\}$ is an arbitrary
orthonormal triad.  The assumption that $\{{\mathbf e}_1, {\mathbf
e}_2, {\mathbf e}_3\}$ can be chosen arbitrarily is justified in 
 Section~\ref{sec:ExtendKS2}.

The fact that there may be errors in the alignment of the
analyzers is already included in the fact that the analyzers
are not assumed  ideal (see
Section~\ref{sec:WhatObs}).  It
may therefore be assumed that the  nominal triad
$\{{\mathbf e}_1, {\mathbf e}_2, {\mathbf e}_3\}$ is strictly
orthogonal.

Suppose that one of these analyzers is used on its own, to measure
the single observable $\bigl( {\mathbf n} \cdot \hat{\mathbf
S}\bigr)^2$.   MKC postulate that the
measurement  will reveal the value of some neighbouring vector
${\mathbf n}' \in S^2_0$ (where, as usual, $S^2_0$ denotes the
dense, KS-colourable set on which the MKC valuations are
defined).  For each ${\mathbf n}$ there is a probability measure
$\mu_{\mathbf n}$ such that $\mu_{\mathbf n} (A)$ is the
probability that ${\mathbf n}' \in A \subseteq
S^2_0$.

Now consider the situation when the analyzers are used jointly, to
measure 
$({\mathbf e}_1 \cdot \hat{\mathbf S})^2$,  
$({\mathbf e}_2 \cdot \hat{\mathbf S})^2$,  
$({\mathbf e}_3 \cdot \hat{\mathbf S})^2$.   Let 
$({\mathbf e}'_1, {\mathbf e}'_2, {\mathbf e}'_3)$ be the triad
whose values are revealed by the measurement.  If the behaviour of
each analyzer is independent of the context in which it is
used, then the probability that 
$({\mathbf e}'_1, {\mathbf e}'_2, {\mathbf e}'_3)
\in A_1 \times A_2 \times A_3$ must be 
$\mu_{{\mathbf e}_1}(A_1) \mu_{{\mathbf e}_2}(A_2) \mu_{{\mathbf
e}_3}(A_3)$. 

It can be seen that, if the behaviour of the analyzers is
context-independent, then the triad 
$\{{\mathbf e}'_1, {\mathbf e}'_2, {\mathbf e}'_3\}$  is not
guaranteed to be strictly orthogonal (unlike the triad 
$\{{\mathbf e}_1, {\mathbf e}_2, {\mathbf e}_3\}$, which is
orthogonal by assumption).  Consequently, the measurement outcome
is not guaranteed to be one of the ``legal'' combinations
$110,101,011$.  In Appendix~\ref{ap:SequentialArgument} we use
this to show that, if the analyzers are sufficiently  near-ideal,
then the assumption of context-independence conflicts with the
empirical predictions of quantum mechanics.   Specifically, we
show that context-independence implies
\begin{equation}
 N \epsilon_{\rm q} \ge \frac{1}{2}
\label{eq:SeqCondA}
\end{equation}
where $\epsilon_{\rm q}$ is the quantum non-ideality index defined
by Eq.~(\ref{eq:EpsQDef}), and $N$ is the smallest number of
distinct orthonormal triads which cannot be consistently
coloured.

Inequality~(\ref{eq:SeqCondA}) shows that, if the analyzers are
sufficiently near-ideal, so that $\epsilon_{\rm q} < 1/(2 N)$,
then the observable whose value is revealed by one analyzer must
depend on the settings of the other two analyzers.  This is a form
of contextuality.

Inequality~(\ref{eq:SeqCondA}) was derived on the assumption that
a measurement of $\bigl( {\mathbf n} \cdot \hat{\mathbf
S}\bigr)^2$ \emph{always} reveals the pre-existing value of
${\mathbf n}' \in S^2_0$.  Suppose we make the weaker assumption,
that the value of ${\mathbf n}'$ is only revealed  with probability
$\ge 1 -\delta_{\rm c}$. 
Then a similar argument to that given in
Appendix~\ref{ap:SequentialArgument} shows 
\begin{equation}
 N \epsilon_{\rm q} + \delta_{\rm c} \ge \frac{1}{2}
\label{eq:SeqCondB}
\end{equation}
if the behaviour of the analyzers is context-independent.  This
inequality should be compared with
Inequality~(\ref{eq:NonIdealKSResult1}).

It should be noted that essentially the same argument goes through
in the case where the measurements are made non-sequentially.   We
have chosen to focus on the sequential case because that is where
the conflict with classical assumptions emerges most clearly.

Kent~\cite{MKC4}
has argued that  sequential measurements are not joint
measurements.  According to him a process does not count as a
joint measurement unless the different observables are measured
``all at once'', in parallel so to speak.  We consider this
objection to be unjustified, for the reasons explained in
Appendix~\ref{ap:KentObjection}.

\section{Are the MKC models Empirically Equivalent to 
Quantum Mechanics?} 
\label{sec:EmpEquiv}
The purpose of this section is to discuss the question, whether
there actually exists a complete theory of MKC type which
is empirically equivalent to quantum mechanics.  In particular, we
will discuss a recent claim by Cabello~\cite{Cabel2}, that
theories of MKC type make empirically testable predictions which
definitely conflict with those of quantum mechanics.  Our
conclusion will be that, although Cabello makes some very
pertinent points, the question remains open. 

To begin with it should be noted that, as Clifton and
Kent~\cite{Kent2} themselves emphasize, the MKC models are, as
they stand, incomplete, since they do not include any dynamical
postulate.
 The  result proved in the last
section illustrates this point.  We showed that, if the
analyzers in Fig.~\ref{fig:sequential} are sufficiently
near-ideal, MKC's assumptions require there to be a delicately
adjusted ``conspiracy'', whereby the behaviour of one analyzer is
influenced by the settings of the other two analzyers in just that
particular manner which is required to secure agreement with the
empirical predictions of quantum mechanics.   In a complete hidden
variables theory this conspiracy would be explained in terms of
the dynamics of the hidden variables describing the
system+apparatus+environment composite.   However, MKC do not
consider the dynamics.  Consequently, their account does not show
that it is  actually  possible to construct a complete, dynamical
theory having all the required properties.

Let ${\mathcal P}_{\rm d}$ be the set of projections on which the
MKC valuations are defined, and let $\Lambda$ be the hidden state
space.  Clifton and Kent~\cite{Kent2} show that one can associate
to each
$\hat{P}
\in {\mathcal P}_{\rm d}$ a function $P\colon \Lambda \to \{0,1\}$,
and to each quantum state $\psi$ a probability measure
$\nu_{\psi}$ such that
\begin{equation}
 \langle \psi | \hat{P} | \psi \rangle = 
\int_{\Lambda} P(\lambda)\, d\nu_{\psi}
\label{eq:CKStatThm}
\end{equation}
Let us call this their \emph{statistical theorem}.

In order to obtain a complete hidden variables theory, they would
need to supplement this result with a \emph{dynamical theorem}. 
They would need to show that there exists a one-parameter family of
mappings
$F_t
\colon
\Lambda \to \Lambda$ such that $F_{t_1} \circ F_{t_2} =
F_{t_1+t_2}$ and
\begin{equation}
\nu_{\psi_t} = \nu_{\psi} \circ F^{-1}_{t}
\label{eq:CKDynamicalThm}
\end{equation}
for all $t$, $\psi$ (where $\psi_t$ denotes the 
Schr\"{o}dinger picture state into which
$\psi$ evolves after a time $t$).   It is a non-trivial question,
whether such mappings actually exist.

There are three  weaknesses in MKC's argument.  MKC's
mathematical results---the existence of dense KS-colourable sets,
and Clifton and Kent's statistical theorem---are clearly
formulated, and rigorously proved.   However, the physical part of
their argument, to show that these results have the effect of 
``nullifying'' the KS theorem, is much less clear, as  
Cabello~\cite{Cabel}  and Mermin~\cite{Mermin1} have stressed. 
The counter-argument we presented in
Sections~\ref{sec:WhatObs}--\ref{sec:NullFurther} was directed at
this point of weakness.  We showed that the physical part of MKC's
argument disintegrates when subjected to careful analysis.

The second point of weakness is the lack of any dynamical
theorem.    If it could be shown that it is definitely not possible
to define a dynamical evolution satisfying
Eq.~(\ref{eq:CKDynamicalThm}), then it would follow that there does
not exist a complete theory of MKC type which is empirically
equivalent to quantum mechanics.  

The third weakness is that MKC only consider joint measurements
of commuting observables.  In a genuinely non-contextual theory
there should be a principle of \emph{approximate}
non-contextuality, applying to joint measurements of
\emph{non-commuting} observables~\cite{me3}.  

The argument in Cabello~\cite{Cabel2}, however, is not directed
at any of these points of weakness.   Instead, Cabello
claims to have derived a result  (his Lemmas~3 and~4) which
directly contradicts Clifton  and Kent's statistical theorem. 
There are only two possibilities here:  either there must be a
technical mistake in the proof of Clifton and Kent's statistical
theorem, or else there must be a technical mistake in the proof of
Cabello's Lemmas~3 and~4.  It appears to us
that the mistake is in fact on Cabello's side.

MKC postulate that the effect of measuring an observable $\hat{P}$
is to reveal the pre-existing value of some other observable
$\hat{P}'$ which is very close to $\hat{P}$.   Cabello (see
footnote~27 in his paper) takes this to imply that a sufficiently
high precision measurement of $\hat{P}$ will, with high
probability, reveal the pre-existing value of $\hat{P}$ itself (if
$\hat{P}$ is in the set ${\mathcal P}_{\rm d}$ on which the
valuation is defined).  In other words he assumes (in the notation
of Eq.~(\ref{eq:CKStatThm}) above) that, if 
$\hat{P}'
\approx
\hat{P}$, then there is a high probability that
$P'(\lambda) = P (\lambda)$. He  fails to allow for the
fact, which we proved in Section~\ref{sec:ExtendKS1} and
Appendix~\ref{ap:Colour}, that the MKC valuations are radically
discontinuous. This means that a knowledge of the
value of
$\hat{P}'$ does not, in general, convey any information at all
regarding the value of $\hat{P}$, no matter how small the
difference  $\hat{P}'-\hat{P}$.

If 
$\hat{P}' \approx\hat{P}$ then it follows from 
Eq.~(\ref{eq:CKStatThm}) that $\int_{\Lambda} \left( P'(\lambda) -
P(\lambda)\right) d\nu_{\psi} \approx 0$ for all $\psi$.  However,
this does \emph{not} imply $\int_{\Lambda} \left| P'(\lambda) -
P(\lambda)\right| d\nu_{\psi} \approx 0$, as assumed by Cabello.

Cabello's argument can, in fact, be turned around, to disprove the
assumption on which it depends.   It follows
from Clifton and Kent's statistical theorem that  the conclusion
to Cabello's Lemma~3 is false.    So what Cabello has really done
is to prove, by
\emph{reductio ad absurdum}, that there are cases where 
$\int_{\Lambda} \left| P'(\lambda) -
P(\lambda)\right| d\nu_{\psi}$ is not negligible, even though
$\hat{P}' \approx\hat{P}$.  He has similarly proved that there
are
 cases for which  there is a non-negligible probability of two
nearly orthogonal projectors both evaluating to 1.

Of course, in a genuinely
non-contextual theory, there \emph{would} be a high probability 
of
$\hat{P}'$ having the same value as
$\hat{P}$, if $\hat{P}'-\hat{P}$ is sufficiently small.  The fact
that the MKC models do not have this  property means that they
are not  non-contextual---as we  showed in  
Section~\ref{sec:ExtendKS1}.  However, Cabello's argument is not
directed at MKC's
 claim, that their models are non-contextual.  It is directed,
instead, at MKC's claim, that their models reproduce the
empirical predictions of quantum mechanics.  There is no 
evident
contradiction between this latter claim and the fact that the
valuations in their models are highly discontinuous.

It should, however, be stressed that, although it has not been
proved that there  does not exist a complete
theory of MKC type which is empirically equivalent to quantum
mechanics, the contrary proposition has not been proved
either.  The question, in short, remains open.  Furthermore,
although Cabello fails to establish his main contention, some of
his intermediate results are of considerable interest.  It is
possible that further investigation along these lines would show
that Cabello's contention is, after all, correct.  However, it
appears to us that such a programme is unlikely to succeed
unless it is directed at one of the points of weakness indicated
above, rather than at Clifton and Kent's statistical theorem
(which seems to have been rigorously proved).

\section{Conclusion}
\label{sec:Conc}

A substantial part of this paper has been devoted to a criticism
of MKC.  We ought to stress, by way of conclusion,
that we actually consider that they have made a very
important contribution to  this subject.
We have argued that MKC's claim, that their models
``nullify'' the KS theorem, was premature; in fact, fallacious.
However, the models themselves are deeply interesting.
They confirm the fact that there are depths to the KS theorem
which, in the past, have not been sufficiently appreciated
(``confirm'' because this point had, in fact, already been
established by the  constructions of
Pitowsky~\cite{Pitowski}).

MKC take it to be almost self-evident that the existence of dense
KS-colourable subsets of
$S^2$ ``nullifies'' the KS theorem.  We have argued that they are
mistaken, and that there is actually no nullification.  
However, the mistake is far from obvious.  MKC's contention
seems, at first sight, highly plausible.  At least, it
seemed highly plausible to the present author, before he began to
seriously engage with the problem.

In order to see that the MKC models do not ``nullify'' the KS
theorem it is necessary to appreciate the point made in
Section~\ref{sec:ExtendKS1}, that a process does not count as a
classical measurement unless it reveals, not simply a pre-existing
\emph{value}, but a pre-existing piece of classical
\emph{information}.  To put it another way, the outcome of a
classical measurement is, not simply a bare
number, but a determinate  proposition, which expresses some
pre-existing classical fact.  It is also necessary to appreciate
the points made in Section~\ref{sec:WhatObs} and 
Appendix~\ref{ap:NonIdeal}, regarding non-ideal quantum
measurements.   In previous accounts these points have not been
explicitly recognized.  They are, however, crucial, as the MKC
models demonstrate.

Although the MKC models do not ``nullify'' the KS theorem, they do
cast important new light on the physical implications of the KS
theorem.  They show that there are subtleties to the physical
interpretation of the KS theorem which have not, in the past, been
sufficiently emphasized.  For this reason the work of MKC is
valuable.   

Although we have presented our discussion in the form of a
criticism it might, in some ways, be more appropriate if it was
seen, in terms which are less prejudicial, as one more
contribution to a line of investigation which
Pitowsky~\cite{Pitowski} initiated, and which MKC have continued. 
It is a line of investigation which might be worth pursuing 
further.

Pitowsky's constructions are particularly interesting in this
respect, because he challenges traditional assumptions at a much
more fundamental level.   For instance, he challenges the 
assumption that sets which are not Lebesgue-measurable have no
physical significance.  Pitowsky's idea, that ``one can, as it
were, reduce physical paradox to a mathematical `pathology' '', is
unlikely to lead to a classical explanation of quantum
phenomena (a ``pathological'' theory---which is to say, a theory
that seems strange as judged by the standards of ordinary
intuition---is a non-classical theory, almost by definition). 
However, that does not seem to be Pitowsky's intention.   Rather,
he seeks to gain additional insight, into what quantum mechanics
is actually telling us,  by considering the problem from a new
and unusual perspective.  It is  a deeply  interesting idea,
which deserves to be  more widely known.

\subsubsection*{Acknowledgments}
It is a pleasure to thank Prof.\ A.~Zeilinger for his hospitality
during the programme ``Quantum Measurement and Information'' held
at ESI in Vienna.  It is also a pleasure to thank 
N.D.~Mermin,  A.~Peres,  A.~Kent,  P.~Busch, D.~Home, K.~Svozil,
I.~Pitowsky, J.~Larsson, G.~Mahler and two anonymous referees for
their stimulating and helpful comments.

\appendix

\section{Approximate von Neumamnn Measurements}
\label{ap:NonIdeal}
In Section~\ref{sec:WhatObs} we criticized MKC's concept, of the
single observable that is \emph{actually} measured, for the special
case of the Stern-Gerlach apparatus.  The purpose of this Appendix
is to show that our argument generalizes, to the case of any
approximate von Neumann measurement (\emph{i.e.} any measurement
described by a unitary operator which is close to the unitary
operator  describing a von Neumann measurement). 

Consider, to begin with, a strict  von Neumann
measurement  of the observable
$\hat{A} =
\sum_{a\in {\mathcal E}}\, a |a\rangle
\langle a|$ (where ${\mathcal E}$ is the set of eigenvalues)
acting on an $n$ dimensional Hilbert space
${\mathcal H}$.  For the sake of simplicity we assume that
$\hat{A}$ is non-degenerate.  In order to perform the measurement
$\hat{A}$ is coupled to a pointer observable $\hat{\mu} = \sum_{a
\in {\mathcal E}}\, a\, |a\rangle_{\rm pt}  \mathstrut_{\rm
pt}\langle a|$ acting on a different $n$-dimensional Hilbert space
${\mathcal H}_{\rm pt}$ (the state space of the measuring
instrument).  The measurement interaction causes the evolution
\begin{equation}
\hat{U} | a \rangle \otimes |\phi_0\rangle_{\rm pt}
= | a \rangle \otimes |a \rangle_{\rm pt}
\label{eq:VonNeu}
\end{equation} 
for all $a$, where $\hat{U}$ is the unitary evolution operator
describing the
 measurement interaction, and $|\phi_0\rangle_{\rm pt}$ is the
(fixed) initial state of the measuring instrument.

The measurement described by Eq.~(\ref{eq:VonNeu}) is ideal
because~\cite{NoteIdealB}
\begin{list}{}{}
\item[1.] It discriminates eigenstates with perfect reliability
(if the system is initially in the eigenstate $|a\rangle$,
then the final pointer reading will certainly be $a$).
\item[2.]  The final state correlations are perfect (if the
final pointer reading is found to be $a$, then the final state of
the system is
 $|a\rangle$).
\end{list}

In practice it would be difficult, if not impossible
to  construct an apparatus such that $\hat{U}$ satisfied
Eq.~(\ref{eq:VonNeu}) \emph{precisely}.  A more realistic
model of the kind of measurement which it might
actually be possible to perform, in the laboratory, would be
\begin{equation}
\hat{U} | a \rangle \otimes |\phi_0\rangle_{\rm pt}
= | a \rangle \otimes |a \rangle_{\rm pt} + \sum_{b,c \in
{\mathcal E}}
\epsilon_{a,bc} \, | b \rangle \otimes |c \rangle_{\rm pt}
\label{eq:ProxVonNeu}
\end{equation} 
where the coefficients $\epsilon_{a,bc}$ are small. 

Of course, the model measurement process described by
Eq.~(\ref{eq:ProxVonNeu}), though more realistic than the one
described by Eq.~(\ref{eq:VonNeu}), is still highly idealized.  A
real laboratory instrument is a complex macroscopic system having 
many other degrees of freedom apart from the pointer observable. 
It is also an open system, undergoing non-negligible
interactions with its environment (these interactions play an
important role in accounting for the fact that system+instrument
are in a mixed state at the end of the measurement
process~\cite{Peres3}).  Lastly, it should be noted that in a real
laboratory measurement the final state correlations are often
(though not always) very poor (for instance, in the traditional
Stern-Gerlach arrangement, where the particle hits a
photographic plate, the result of the measurement  does not
contain useful information regarding the
\emph{final} spin state of the particle). However, the model is
sufficiently realistic for present purposes.   

The unitary operator $\hat{U}$ in 
Eq.~(\ref{eq:ProxVonNeu}) is close to the unitary operator
describing a von Neumann measurement.  In other words, it
describes an
approximate von Neumann measurement. The process is not an
ideal measurement of
$\hat{A}$.  It is, however,  a \emph{non-ideal}
measurement (which is the most that can reasonably be demanded of
a real laboratory procedure).  Specifically:
\begin{list}{}{}
\item[$1'$.] It discriminates eigenstates with a high degree of
reliability (if the system is initially in the eigenstate
$|a\rangle$, then there is a high probability
 that the final
pointer reading will approximately be
$a$).
\item[$2'$.]  The final state correlations are good (if the
final pointer reading is found to be $a$ then ${\rm
Tr}\bigl((\hat{A} - a)^2 \hat{\rho}
\bigr)\approx 0$, where $\hat{\rho}$ is the reduced density
matrix describing the final state of the system, after the
pointer has been read).
\end{list}
These properties should be compared with the defining
characteristics of an ideal measurement  (items~1 and~2 
above).

We now show that the process also effects a non-ideal measurement
of every other observable which is sufficiently close to 
$\hat{A}$.

Choose some continuous parameterization for the ONBs (orthonormal
bases) in
${\mathcal H}$, so that $|a,\nu\rangle$ is the ONB with parameter
value
$\nu$.  Let
$\nu_{0}$ be parameter value corresponding to the original ONB: 
$|a, \nu_{0}\rangle = |a\rangle$.
Then an arbitrary observable $\hat{B}$ can be expressed in the
form 
\begin{equation}
\hat{B} = \hat{A}(\nu, \bbox{\delta})
=\sum_{a \in {\mathcal E}} (a+\delta_a) |a,\nu\rangle\langle a,
\nu|
\end{equation}
for suitable $\nu, \bbox{\delta}$.  If  $\hat{B}$ is close
to $\hat{A}$ then we can choose 
$\nu
\approx
\nu_0$ and
$\bbox{\delta} \approx 0$.  

Eq.~(\ref{eq:ProxVonNeu}) can be re-written in terms of the ONB
$|a,\nu\rangle$:
\begin{equation}
\hat{U} | a,\nu \rangle \otimes |\phi_0\rangle_{\rm pt}
= | a,\nu \rangle \otimes |a \rangle_{\rm pt} + \sum_{b,c \in
{\mathcal E}}
\epsilon_{a,bc}(\nu) \, | b, \nu\rangle \otimes |c \rangle_{\rm pt}
\end{equation} 
where the coefficients $\epsilon_{a,bc}(\nu)$ vary continuously
with $\nu$.  It follows that, if $\nu \approx \nu_0$ and
$\bbox{\delta} \approx 0$, then the coefficients
$\epsilon_{a,bc}(\nu)$ will  be small, and so  the process will
be a non-ideal measurement of
$\hat{A}(\nu, \bbox{\delta})$ (in the sense of the definition
expressed by items~$1'$ and~$2'$ above).

We conclude that $\hat{U}$ describes a non-ideal measurement of
$\hat{B}$ for every $\hat{B}$ close to $\hat{A}$---as claimed.

MKC assume that, given any non-ideal measurement  of $\hat{A}$,
there will be a  neighbouring observable 
$\hat{B}$ which is measured ideally, and which they take to be the
only observable that is ``actually'' measured.
We noted in Section~\ref{sec:WhatObs} that there exists at least
one class of non-ideal measurements  for which this assumption is
invalid:  namely, Stern-Gerlach measurements.  A
Stern-Gerlach measurement is not an ideal measurement of
\emph{anything}:  not an ideal measurement of the nominal
observable, which the experimeter records as having been measured,
and not an ideal measurement of any other observable
either~\cite{BuschBk}. Measurements of this kind might be
described as
\emph{completely} non-ideal.   It is easily seen that they are, in
a certain well-defined sense, typical, or generic.

If the measurement  described by Eq.~(\ref{eq:ProxVonNeu}) is of
the special kind assumed by MKC then
\begin{equation}
\hat{U} | a,\nu \rangle \otimes |\phi_0\rangle_{\rm pt}
= e^{i \theta_a} | a,\nu \rangle \otimes |a \rangle_{\rm pt}
\end{equation} 
for some $\nu\approx\nu_0$ and vector $\bbox{\theta}$. We
may thus identify the set of all such measurements with a
neighbourhood of the identity in the group  
$U(n)$, having dimension $n^2$.  On the other hand, the space of
coefficients
$\epsilon_{a,bc}$ may be identified with  a neighbourhood in the
coset space $U(n^2)/U(n^2-n)$, having dimension $n^2(2n-1)$.   It
follows that, if $n>1$, the set of measurements assumed by MKC
constitute a proper, lower-dimensional submanifold of the manifold
of all non-ideal measurements.  This means that, if one were to
select a set of coefficients $\epsilon_{a,bc}$ satisfying the
unitarity condition at random, then, for many  reasonable
choices of probability measure, there would be probability zero
that the measurement which resulted was of the special kind
assumed by MKC. In any case, the contrary  hypothesis, that a
randomly chosen laboratory measurement is likely to be of the kind
assumed by MKC, cannot be considered plausible.

Suppose, however, that a laboratory measurement of $\hat{A}$ did
happen to be of the very special kind assumed by MKC.  Then there
would be some observable $\hat{B}\approx\hat{A}$ which was
measured ideally.  However, MKC would still not be justified in
arguing that $\hat{B}$ is the only observable that is ``actually''
measured.  The fact that the measurement of  $\hat{B}$ is slightly
better than the measurement  of $\hat{A}$ does not imply that
$\hat{A}$ is not measured \emph{at all}.  A measurement does not
have to be strictly ideal in order to count as a measurement
(otherwise one would have to say that  a Stern-Gerlach apparatus
does \emph{not} measure spin, and that real photon detectors do
\emph{not} detect photons).

\section{Lemmas Concerning the Regions $C$ and $D$.}
\label{ap:Colour}
The purpose of this appendix is to prove those results to
which we appealed, without proof, in Section~\ref{sec:ExtendKS1}. 
For ease of reference we state the results as
lemmas.   The reader  should refer to Section~\ref{sec:ExtendKS1}
for definitions of notation and terminology.

\begin{quote}
{\bfseries Lemma 1}\ \  \emph{The continuity region $C$ is
KS-colourable.  Furthermore $f$  extends by continuity to a
uniquely defined continuous KS-colouring
$\bar{f}\colon  C \to
\{0,1\}$ }.
\end{quote}

Let  $C_0$ 
(respectively $C_1$) consist of those vectors 
${\mathbf n} \in C$ having a neighbourhood $U$
such that $f({\mathbf n}') = 0$ (respectively 
$f({\mathbf n}') = 1$) for all ${\mathbf
n}'
\in  U\cap S^2_0$.  

$C_0$ and $C_1$ partition $C$ into two disjoint
(open)
 subsets.  We may therefore define
  a function $\bar{f} \colon C\to\{0, 1\}$ by setting
$\bar{f}({\mathbf n}) = 0$ if ${\mathbf n} \in C_0$, and
$\tilde{f}({\mathbf n}) = 1$ if ${\mathbf n} \in C_1$.  

Now 
let $\{ {\mathbf e}_1, {\mathbf e}_2, {\mathbf e}_3\}$ be any
orthonormal
triad $\subseteq C$.  By construction, there exists a triad
$\{ {\mathbf e}'_1, {\mathbf e}'_2, {\mathbf e}'_3\} \subseteq
C\cap S^2_0$ such that $\bar{f}({\mathbf e}_r) =f({\mathbf
e}'_r)$ for
$r=1,2,3$.  It follows that   $\{ {\mathbf e}_1, {\mathbf e}_2,
{\mathbf e}_3\}$ must $\bar{f}$-evaluate to one of the
``legal'' combinations $011, 101, 110$.  A similar argument shows
that two orthogonal vectors $\in C$ cannot both
$\bar{f}$-evaluate to 0, even if their vector product 
 $\notin C$.  

It follows that $\bar{f}$ is a KS-colouring of
the set $C$.  The claim is now immediate.

\begin{quote}
{\bfseries Lemma 2}\ \  \emph{The discontinuity region $D$ has
non-empty interior.}
\end{quote}

Let $\{{\mathbf n}_1, {\mathbf n}_2, \dots
{\mathbf n}_R \}$ be any  finite KS-uncolourable set.  The
fact that 
$C$ (the complement of $D$) is KS-colourable means that at least
one of these vectors must $\in D$.  Suppose that the
labelling is such that, for some $r\ge 1$, 
$\{{\mathbf n}_1, {\mathbf n}_2, \dots
{\mathbf n}_r \} \subseteq D$ and 
$\{{\mathbf n}_{r+1}, {\mathbf n}_{r+2}, \dots
{\mathbf n}_R \} \subseteq C$. 

For each $s = 1, \dots, r$ let $\epsilon_{s}$ be the angular
distance from ${\mathbf n}_s$ to $C$, and for  each
$t = r+1, \dots, R$ let $\delta_{t}$ be the angular distance
from ${\mathbf n}_t$ to $D$.  Define $\delta =
\min_{(r+1)\le t \le R} \bigl(\delta_t\bigr)$.  The fact that
$C$ is open means that $\delta >0$ (unless $C$ is empty---in which
case the claim is immediate). 

If it should happen that $\epsilon_s >0$ for some $s$, then 
${\mathbf n}_s$ is in the interior of $D$---which
means that the interior of $D$ is non-empty. 
Otherwise we can find a suitable rotation,  through an angle
$<
\delta$, which will take one or more of the vectors  ${\mathbf
n}_1,
\dots, {\mathbf n}_r$
into
$C$, while not taking any of the vectors ${\mathbf n}_{r+1},
\dots, {\mathbf n}_R$ out of
$C$.  After re-labelling this gives us a new KS-uncolourable
set
$\{{\mathbf n}'_1, {\mathbf n}'_2, \dots
{\mathbf n}'_R \}$ with
 $\{{\mathbf n}'_1, {\mathbf n}'_2, \dots
{\mathbf n}'_{r'} \} \subseteq D$ and 
$\{{\mathbf n}'_{r'+1}, {\mathbf n}'_{r'+2}, \dots
{\mathbf n}'_R \} \subseteq C$ for some $r' <r$.

If it should still happen that none of the vectors
${\mathbf
n}'_1,
\dots, {\mathbf n}'_{r'}$ is in the interior
of $D$ we can repeat the procedure.  

It is impossible to
rotate all the vectors into $C$, so after sufficiently many
iterations of the procedure at least one of the vectors must
be in the interior of
$D$.  It follows that the interior of $D$
cannot be empty.

We conclude by noting that this argument actually establishes the
 more general statement, that the
complement of \emph{any} open KS-colourable set must have non-empty
interior.

\begin{quote}
{\bfseries Lemma 3}\ \  \emph{There is a fixed, model-independent,
non-zero lower bound on the solid angle subtended by $D$}
\end{quote}

Let us note that it already follows from Lemma~2 that the solid
angle  subtended by $D$ must be $>0$, for any \emph{given}
colouring.  Lemma~3
implies a stronger statement: that one cannot, by choosing the
colouring appropriately, make the solid angle arbitrarily small.

We begin by establishing some notation. Let $\mu$ be the invariant
measure on $S^2$, normalized so that $\mu(S^2)=1$ 
(\emph{i.e.}\ the solid angle scaled by  $1/(4 \pi)$).
Let ${\mathcal O}$ (respectively ${\mathcal B}$) be the  set of
all open (respectively Borel, or $\mu$-measurable) subsets of
$S^2$ which are KS-colourable.  Define
\begin{eqnarray}
d_{\mathcal O}& = & 1 - \sup_{U \in \mathcal O} \bigl(
\mu(U)\bigr)\\
d_{\mathcal B}& = & 1 - \sup_{B \in \mathcal B} \bigl(
\mu(B)\bigr)
\end{eqnarray}
Clearly, $d_{\mathcal O} \ge d_{\mathcal B}$ (it is a non-trivial
question as to whether $d_{\mathcal O}$ actually $ = d_{\mathcal
B}$).

We will show that $d_{\mathcal B}>0$.  In view of the fact that $C
\in {\mathcal O}$ and $D= S^2 - C$ this implies
\begin{equation}
\mu(D)  \ge d_{\mathcal O} \ge d_{\mathcal B} >0
\end{equation}
which  establishes the claim.

Before proceeding to the proof, let us note that it follows from
the last paragraph in the proof of Lemma~2 that $\mu(U) >0$ for
all $U \in {\mathcal O}$.  The proof which follows is needed to
show that there is not a sequence of sets 
$U_n \in {\mathcal O}$ such that $\mu(U_n ) \to  0$.  This may
appear intuitively obvious.  However, the fact that there exist
dense KS-colourable subsets of $S^2$ was intuitively surprising (at
least to the present writer), and so it behoves us to be careful.

We now proceed to the proof that  $d_{\mathcal B} >0$.

Choose some fixed, finite,  KS-uncolourable set  $\{{\mathbf
n}_1, {\mathbf n}_2, \dots , {\mathbf n}_{2 M}
\} \subset S^2$, with the property that 
${\mathbf n}_{M+r} = -{\mathbf n}_r$ for $r=1,\dots, M$.

Let
$\theta_0$ be the minimum angular separation of the vectors
belonging to this set:
\begin{equation}
 \theta_{0} = \min_{1\le r, s \le M}
\bigl( \cos^{-1} ({\mathbf  n}_r \cdot {\mathbf n}_s) \bigr)
\end{equation}

Surround each vector ${\mathbf n}_r$ with a circular patch
$E_r$ of radius $\theta_0/2$:
\begin{equation}
 E_r = \{ {\mathbf m} \in S^2 \colon\  \cos^{-1} ({\mathbf m}
\cdot {\mathbf n}_r) \le \theta_0 /2\}
\end{equation}

Let  $B$ be any set $\in {\mathcal B}$, and let $B^{\rm c}
=S^2- B$ be its complement.  We may assume, without loss of
generality, that $B$ is invariant under the parity operation
(implying that $B^{\rm c}$ is also invariant).  By
construction, the sets
$E_r$ are non-intersecting, with the possible exception of a set of
measure zero on their boundaries.  Consequently
\begin{equation}
 \mu (B^{\rm c}) \ge \sum_{r=1}^{2M} \mu (E_r \cap B^{\rm c})
\label{eq:BCompEst1}
\end{equation}

We next define, for each $r$, a function $g_r \colon 
S^2 \to E_r$ by
\begin{equation}
 g_r( {\mathbf m}) = e^{(\theta_0/2) {\mathbf m} \cdot {\mathbf
L}} {\mathbf n}_r
\end{equation}
where $L_1, L_2, L_3$ are the generators of ${\rm SO} (3)$. 
Thus, $g_r({\mathbf m})$ is the vector  obtained
by rotating ${\mathbf n}_r$ through a (fixed) angle $\theta_0/2$
about the (variable) axis ${\mathbf m}$.  It is easy to see that,
as
${\mathbf m}$ ranges over the whole of $S^2$, the
interior of $E_r$ is covered  twice, and the boundary once. 
Consequently
\begin{equation}
  \mu (E_r \cap B^{\rm c})
= \frac{1}{2} \int_{\tilde{B}_r^{\rm c}}
J_r ({\mathbf m}) \, d \mu
\label{eq:ECapBC}
\end{equation}
where $J_r$ is the Jacobian of $g_r$, and
$\tilde{B}_r^{\rm c} =g_r^{-1} (E_r \cap B^{\rm c}) $.
Define, for each ${\mathbf m} \in S^2$, 
\begin{equation}
 J ({\mathbf m}) = \min_{1\le r \le 2 M} \bigl(J_r ({\mathbf m}) 
\label{eq:JDef}
\bigr)
\end{equation}
Eqs.~(\ref{eq:BCompEst1})
and~(\ref{eq:ECapBC}) then imply
\begin{equation}
\mu (B^{\rm c})
\ge \frac{1}{2} \left(\int_{\cup_{r=1}^{ M}\tilde{B}_r^{\rm
c}} J ({\mathbf m})\, d \mu + 
\int_{\cup_{r=M+1}^{2 M}\tilde{B}_r^{\rm
c}} J ({\mathbf m})\, d \mu
\right)
\end{equation}
We now observe that, for each fixed ${\mathbf m} \in S^2$,  the
set
$\{ g_1 ({\mathbf m}), g_2 ({\mathbf m}), \dots , g_{2M} ({\mathbf
m})
\}$, being obtained by rotating the KS-uncolourable
set $\{{\mathbf n}_1, {\mathbf n}_2, \dots
{\mathbf n}_{2M} \}$, must itself be KS-uncolourable.   On the
other hand, $B$ is KS-colourable.  It follows that, for each
${\mathbf m} \in S^2$, we must have $g_r ({\mathbf
m})
\in B^{\rm c} $ and 
$g_{M+r} ({\mathbf
m})
\in B^{\rm c} $ for some $r\le M$.  Consequently
$\cup_{r=1}^{M} \tilde{B}_r^{\rm c} =\cup_{r=M+1}^{2M}
\tilde{B}_r^{\rm c}= S^2$, and 
\begin{equation}
 \mu (B^{\rm c}) \ge  \int J({\mathbf m})\, d \mu
\end{equation}
Since the  right-hand side of this inequality is 
independent of $B$  we may deduce
\begin{equation}
 \sup_{B \in {\mathcal B}} \bigl(\mu (B )\bigr)
\le 1 -  \int J({\mathbf m})\, d \mu
\label{eq:SupBBound}
\end{equation}
To complete the proof we note that $J({\mathbf m})$ is a
continuous, non-negative function having finitely many (in fact
$2 M$) zeros.  It follows that 
\begin{equation}
d_{\mathcal B} \ge \int J({\mathbf m}) \, d
\mu >0
\label{eq:dcBound}
\end{equation}
\subsection*{Concluding  Remarks}
It is natural  to consider, in addition to the  sets ${\mathcal
O}$ and ${\mathcal B}$ defined above, the set ${\mathcal C}$,
consisting of all closed, KS-colourable subsets of $S^2$.  Define
$d_{\mathcal C} = 1 - \sup_{K \in {\mathcal C}} \bigl(
\mu(K) \bigr)$.  Clearly, 
$d_{\mathcal C} \ge d_{\mathcal B}$.   It is  a standard result
(see, for example, Halmos~\cite{Halmos}, Chap.~10) that, for each
Borel set $B$, there exists a sequence of closed sets $K_n
\subseteq B$ such that $\mu(B) = \lim_{n \to \infty} \bigl(
\mu(K_n)
\bigr)$.  It follows that we also have $d_{\mathcal C} \le
d_{\mathcal B}$.  Combining these inequalities we deduce
\begin{equation}
d_{\mathcal O} \ge d_{\mathcal C} = d_{\mathcal B}
\end{equation}
 
The numbers $d_{\mathcal O}$ and $d_{\mathcal C} = d_{\mathcal B}$
could be regarded as non-classicality indices (the larger they are,
the more radically quantum mechanics departs from classical
assumptions).  From this point of view it would be interesting to
know their actual values.

The evaluation of the integral on the right-hand side of
Inequality~(\ref{eq:dcBound}), though straightforward, would be
somewhat tedious.  We will therefore content ourselves with noting
that it follows from Eq.~(\ref{eq:JDef}) that
\begin{equation}
  \int J({\mathbf m})\, d \mu
<  \int J_r ({\mathbf m}) \, d \mu
= 2 \int_{E_r}  d \mu
= 2 \sin ^2 \left(\frac{\theta_0}{4}\right)
\end{equation}
where $r$ is any integer in the range $1, \dots , 2M$, and where
$\theta_0$ is the minimum angular separation of the vectors in the
KS-uncolourable set considered.  In the case of the Conway-Kochen
set~\cite{PeresBk,BubBk} one has $\theta_0 = 18.4^{\rm o}$ (the
angle between the vectors $(0,1,2)$ and $(0,2,2)$), implying that
the right-hand side of Inequality~(\ref{eq:dcBound}) is $< 0.013$.

One would like to know if  $d_{\mathcal O}$ is of the
same order  as the lower bound set by
Inequality~(\ref{eq:dcBound})---implying that 
$\gtrsim 99\%$ of the area of
$S^2$ can be covered with an open KS-colourable subset---or whether
the actual value is significantly  larger.   This is a question
which it might be interesting to investigate further.
It might also be interesting to investigate the analogous
quantities defined on Hilbert spaces of dimension $>3$.  In
particular, it might be interesting to investigate the way in
which these quantities vary with the dimension of the Hilbert
space.

\section{The ``Operational'' Approach of Simon \emph{\bfseries et
al} and Larsson}
\label{ap:Zeilinger}
Simon \emph{et al}~\cite{Zeil1,Simon2} and Larsson~\cite{Lars}
(SBZL) claim to have proved an operational generalization of the KS
theorem, applying to real experiments (Larsson, however, does not 
use the word ``operational'').  Their result inspired the 
generalized KS theorem proved in Section~\ref{sec:ExtendKS2} of
this paper.  It also inspired part of the argument in
Section~\ref{sec:Seq}. However, although we are very significantly
indebted to SBZL, we also have some  significant criticisms.  We
will argue, in fact,  that their operational version of the
theorem is invalid.

In the following we
begin by criticizing SBZL's operational definition of
contextuality.  We go on to identify a loophole in SBZL's
argument.  We conclude by exhibiting a model which gets through
the loophole, and which  provides a concrete counter-example to
their operational  version of the generalized KS theorem.

SBZL consider an instrument which performs  measurements
of each of the $N$ triads contained in some fixed KS uncolourable
set.  The triad to be measured is determined by three switches. 
They assume that the  system+instrument  combination is fully
specified by:  (1)  the  three  switch positions
${\mathbf e}_1,{\mathbf e}_2,{\mathbf e}_3$, (2) the system
hidden variables
$\lambda_{\rm S}$, (3) the  instrument hidden
variables $\lambda_{\rm I}$ (Larsson, however,
denotes the pair $(\lambda_{\rm S}, \lambda_{\rm I})$ by the single
symbol $\lambda$).  In general, the $r^{\rm th}$
pointer reading is given by  a function 
$X_r ({\mathbf e}_1, {\mathbf e}_2, {\mathbf e}_3, \lambda_{
\rm S}, \lambda_{\rm I})$ which depends on all three switch
positions, as well as the hidden variables.  SBZL take it that a
theory is non-contextual in an operational sense  if and only if
there exists, for each measuring instrument,  a single function
$X$ such that
\begin{equation}
X_r ({\mathbf e}_1, {\mathbf e}_2, {\mathbf e}_3, \lambda_{\rm
S}, \lambda_{\rm I})
=
X({\mathbf e}_r, \lambda_{\rm S}, \lambda_{\rm I})
\label{eq:SBZLNonContextDef}
\end{equation}
for $r=1,2,3$.   

This definition has the following features:  (1) if the instrument
is different then the function
$X$ may be different and (2)
$X$ depends on the instrument hidden variables $\lambda_{\rm I}$,
as well as the system hidden variables $\lambda_{\rm S}$.   For
these reasons it is not a definition of non-contextuality in the
ordinary sense.

A non-contextual theory (in the ordinary sense) is a theory in
which measurements reveal the pre-existing values of the
observables measured.  It is, in other words, a classical theory,
at least so far as the concept of measurement is concerned. 
Non-contextuality in  SBZL's operational sense, by contrast, does
not have the implication that measurements reveal pre-existing
values (of the observables measured).  It is not a
 signature of classicality.  In fact, we will show that it is
not even inconsistent with the empirical predictions of quantum
mechanics.  It is therefore inappropriate.  At least, it is
inappropriate if the aim is to identify hypotheses which
the empirically verifiable predictions of quantum mechanics 
exclude.

Before proceeding further it will be convenient to digress, and
consider the ordinary concept of non-contextuality in a little
more detail.

The context of a measurement   is the
particular procedure used to carry out the measurement. 
 Alternatively, it is what Bohr~\cite{Bohr} describes as the
``whole experimental arrangement'', and what Bell~\cite{Bell1}
describes as the ``complete disposition of the apparatus''.  
The term ``contextuality'' or ``context-dependence'' refers to the
phenomenon in which different procedures for  measuring the same
observable have different outcomes, even though the hidden state
of the system is initially the same.

A theory in which measurements reveal pre-existing values (of the
observables measured) is obviously guaranteed to be
non-contextual.  The converse is also true: if a
theory is non-contextual, then  measurements  reveal pre-existing
values (of the observables measured).   It may perhaps happen that
pre-existing values are not explicitly mentioned, in the formal
description of a non-contextual theory.  Nevertheless, they must
be present implicitly (since, if the theory predicts that  a
measurement of
$\hat{A}$ will always produce the value
$a$, then that effectively means the theory assigns to $\hat{A}$
the value
$a$).
To say that a theory is non-contextual is 
\emph{logically equivalent} to
saying that it is a theory in which measurements reveal
pre-existing values (of the observables measured).

There is a
tendency, in formal discussions, to focus on one particular kind of
contextuality:  namely, the kind where the result of measuring
$\hat{A}$  depends on which other commuting observables $\hat{B},
\hat{C}, \dots$ are jointly measured along with
$\hat{A}$.   It is an important kind of contextuality because it
lends itself to the   proof of general theorems.  However, there
are many other kinds.

A difference in measurement context is \emph{any} difference in
the detailed specification of the apparatus used for performing
the measurement.  It need not be a major difference.
Suppose, for example, that the shape
of the pole pieces in  a Stern-Gerlach arrangement was slightly
changed.  This would count as a change of  context.  A real
laboratory instrument admits numerous minor variations on the
basic design.  A theory is not genuinely non-contextual unless
measurement outcomes are insensitive to every such variation.
 
In
Eq.~(\ref{eq:SBZLNonContextDef}) the function $X$ depends on the
instrument, and on
$\lambda_{\rm I}$.  It follows that SBZL's operational criterion
does not imply that measurement outcomes are genuinely independent
of context.  SBZL tend to focus on the switch positions and pointer
readings as the features of the measuring apparatus of which the
experimenter has knowledge (``the switch position is all he knows
about'', as it is expressed in ref.~\cite{Zeil1}).  However a real
measuring apparatus  has many other controllable parameters:  for
example, the number of turns in each coil, the exact position of
each screw, the pressure in each evacuated chamber, \emph{etc.}\
\emph{etc}.  A small change in one of these parameters will be
reflected by a change in the probability distribution on the
variables $\lambda_{\rm I}$, if not by a change in the
function $X$ itself.  Consequently, it may have a large affect on
the measurement outcome, even in a theory satisfying SBZL's
operational criterion.

SBZL's operational criterion is, in one way,
too weak.  Not only does their criterion fail to exclude models
which are highly non-classical. It is not even inconsistent with
the empirical predictions of quantum mechanics (see the
example  at the end of this appendix).

Their criterion is, in another way, too strong.  We
stated earlier that a non-contextual theory is one in which
measurements reveal pre-existing values (of the observables
measured).  However, this only applies to ideal measurements. 
Non-ideal measurements do not certainly  reveal exact
pre-existing values, and so they may only be approximately
independent of context.  SBZL's criterion is meant to apply to
real laboratory measurements.  It is therefore unduly restrictive
in requiring that, for given
$\lambda_{\rm S}$,
$\lambda_{\rm I}$,     
measurements of
${\mathbf e}_1$ in the contexts 
$\{{\mathbf e}_1, {\mathbf e}_2, {\mathbf e}_3\}$ and   
$\{{\mathbf e}_1, {\mathbf e}'_2, {\mathbf e}'_3\}$ must,
\emph{with certainty},
 have the same outcome (this point is closely related to the
point we made at the end of Section~\ref{sec:ExtendKS2}, when we
argued that the condition $N
\epsilon_{\rm q} <1$ is not sufficient to ensure a contradiction
with classical assumptions).

In short, the criterion for a theory to be genuinely
non-contextual is, not Eq.~(\ref{eq:SBZLNonContextDef}), but
\begin{equation}
X_r ({\mathbf e}_1, {\mathbf e}_2, {\mathbf e}_3, \lambda_{\rm
S}, \lambda_{\rm I})
=
f({\mathbf e}_r, \lambda_{\rm S})
\label{eq:SBZLNonContextDefB}
\end{equation}
for all $\lambda_{\rm S}, \lambda_{\rm I}$ in a set of probability
measure $\approx 1$, and for some function $f$ which (unlike
$X$ and $X_r$) is the same for all measuring instruments.   It will
be seen that this is, in effect,  the criterion we used in
Section~\ref{sec:ExtendKS2}.

Let us now consider SBZL's argument.  

SBZL  consider a set of $N$ 
switch settings  
$({\mathbf e}^{1}_{1},{\mathbf e}^{1}_{2},{\mathbf
e}^{1}_{3}), \dots,
({\mathbf e}^{N}_{1},{\mathbf e}^{N}_{2},{\mathbf
e}^{N}_{3})$ which cannot  consistently be coloured. For each
$r$, let $A_r$ be
the set of hidden variables $(\lambda_{ S}, \lambda_{ I})$
such that
\begin{equation}
 X({\mathbf e}^{r}_{1},\lambda_{ S}, \lambda_{ I})
 +
 X({\mathbf e}^{r}_{2},\lambda_{ S}, \lambda_{ I})+
 X({\mathbf e}^{r}_{3},\lambda_{ S}, \lambda_{ I})
 \neq 2
\end{equation}
where $X$ is the
function defined by Eq.~(\ref{eq:SBZLNonContextDef}) above.

Let $\Lambda_{\rm S}$ and $\Lambda_{\rm I}$ be the hidden state
spaces of system and instrument respectively, and let $\mu$
be the probability measure on $\Lambda_{ S}\times
\Lambda_{ I}$ describing the
system+instrument combination.
SBZL  make the crucial assumption, that
$\mu$ is  the same for all switch settings. 

It follows from the definitions that
$\mu(A_r)
\le
\epsilon_{\rm q}$ for all $r$ (where $\epsilon_{\rm q}$ is the
quantity which SBZL denote $\epsilon$, and which is defined by
Eq.~(\ref{eq:EpsQDef}) of this paper).  In view of the fact
that $\cup_{r=1}^{N} A_r = \Lambda_{ S} \times \Lambda_{ I}$
this implies
\begin{equation}
 1 = \mu\left(\bigcup_{r=1}^{N}  A_r \right)
 \le \sum_{r=1}^{N} \mu (A_r) \le N \epsilon_{\rm q}
\end{equation}
SBZL conclude that theories which are non-contextual in their
operational sense cannot reproduce the quantum
mechanical predictions for an apparatus with
$\epsilon_{\rm q}<1/N$.

The problem with this argument is the fact that it rests on the
assumption that $\mu$ is the same for all switch settings.  The
assumption is not justified.  Changing the switch settings
changes the quantum state of the instrument.  In a hidden
variables theory   the quantum state of the instrument is
represented by a probability measure on $\Sigma\times \Lambda_{\rm
I}$ (where $\Sigma$ is the space of switch settings).  It is
reasonable to assume that this probability measure will be of the
form $\delta \times \nu$, where $\delta$ is a $\delta$-function
measure on $\Sigma$.  However, there is no reason to assume that
$\nu$ is the same for all switch settings---which is what would be
needed if SBZL's assumption was to be justified. 

To find an example of a hidden variables theory to which SBZL's
assumption is inapplicable, one does not need to look further than
the Bohm theory.  In the Bohm theory the probability distribution
is given by the square of the modulus of the wave function, in the
$x$-space representation.  Changing the switch settings  changes
the wave function, and so it will, in general, change the
probability distribution.    

It is not simply that SBZL's argument is not sufficient to
establish their conclusion.   Their  conclusion is not true---as we
now show, by means of a counter-example.  The example we choose is
a model of MKC type. It has the property  
$\epsilon_{\rm q}=0$.  The model is not non-contextual in the
ordinary sense, defined by Eq.~(\ref{eq:SBZLNonContextDefB}). 
However, it \emph{is} 
non-contextual in SBZL's operational sense, defined
by Eq.~(\ref{eq:SBZLNonContextDef}).

Let $S^2$ be the unit 2-sphere, and let $S^2_0$ be the dense,
KS-colourable subset on which the MKC valuations are defined.  For
each
$\lambda_{\rm S}$, let $f_{\lambda_{\rm S}}\colon S^2_0 \to
\{0,1\}$ be the associated valuation. 

If the hidden state of the instrument is $\lambda_{\rm I}$ the
measurement reveals the pre-existing values of a triad
$\tau_{\rm I} (\lambda_{\rm I})=
\{{\mathbf e}_{{\rm I} 1}(\lambda_{\rm I}),{\mathbf
e}_{{\rm I} 2}(\lambda_{\rm I}),{\mathbf e}_{{\rm I}
3}(\lambda_{\rm I})\}\subseteq S^2_0$.   The probability
measure on $\Lambda_{\rm I}$ is such that  $\tau_{\rm I}
(\lambda_{\rm I})$ is always extremely close to the triad of
switch settings (this means that the probability measure
must be strongly dependent on the switch settings---contrary to
what is assumed by SBZL).

Given any vector ${\mathbf n} \in S^2$  let
$s({\mathbf n},
\lambda_{\rm I})$ be the index $s$ for which 
the scalar product ${\mathbf e}_{{\rm I}s}(\lambda_{\rm I}) \cdot
{\mathbf n}$ is maximal (ambiguities being resolved by taking the
smallest such index). Then the $r^{\rm th}$ pointer reading is
given by
\begin{equation}
X_r ({\mathbf e}_1, {\mathbf e}_2, {\mathbf e}_3, \lambda_{\rm S},
\lambda_{\rm I}) = f_{\lambda_{\rm S}} \left( {\mathbf
e}_{{\rm I}s({\mathbf e}_r, \lambda_{\rm I})}
(\lambda_{\rm I} )
\right)
\end{equation}
Comparing this equation with Eq.~(\ref{eq:SBZLNonContextDef}) it
can  be seen that the model is non-contextual in 
SBZL's operational sense
(with $X({\mathbf n}, \lambda_{\rm S}, \lambda_{\rm I}) = 
f_{\lambda_{\rm S}} \left( {\mathbf
e}_{{\rm I}s({\mathbf n}, \lambda_{\rm I})}
(\lambda_{\rm I} )
\right)$).  On
the other hand the fact that
$f_{\lambda_{\rm S}}$ is a KS-colouring implies $\epsilon_{\rm
q} =0$.

The example thus shows that SBZL's operational version of the
generalized KS theorem is invalid.  

Of course, the model is
non-contextual in the ordinary sense, as defined by
Eq.~(\ref{eq:SBZLNonContextDefB}).  Consequently, there is no
contradiction with the result proved in
Section~\ref{sec:ExtendKS2}:  that  theories which are
non-contextual in the ordinary  sense cannot reproduce the quantum
mechanical predictions for an apparatus with
$\epsilon_{\rm q} \ll 1/N$.

The model actually shows that non-contextuality in  SBZL's
operational sense is not inconsistent with the empirical
predictions of quantum mechanics, even as regards the outcome of
measurements for which $\epsilon_{\rm q}=0$.  Consequently, it also
provides some additional grounds for considering their operational
criterion to be inappropriate.

\section{Contextuality of Sequential Analyzers}
\label{ap:SequentialArgument}

The purpose of this appendix is to prove
Inequality~(\ref{eq:SeqCondA}), establishing that the observable
whose value is revealed by  a near-ideal analyzer  must depend on
context.

Let $f_{\lambda}$ be the KS-colouring of $S^2_0$ defined by the
hidden state $\lambda$.  Define an induced valuation
$\tilde{f}_\lambda$ of the \emph{full} unit 2-sphere $S^2$ by
\begin{equation}
\tilde{f}_{\lambda} ({\mathbf n})
=
\left\{
\begin{array}{ll}
0 &\hspace{0.5 in} \text{if 
$\mu_{\mathbf n} (\{{\mathbf n}'\in S^2_0: \;
f_{\lambda}({\mathbf n}') =1\})<0.5$}
\\ 1 & \hspace{0.5 in} \text{if
$\mu_{\mathbf n} (\{{\mathbf n}'\in S^2_0: \;
f_{\lambda}({\mathbf n}') =1\})\ge0.5$}
\end{array}
\right.
\end{equation}
for all $\mathbf n$.  
The significance of this construction is
that, if the hidden state is $\lambda$, then there is probability
$\ge 0.5$ that a measurement of 
${\mathbf n}$ will have outcome  
$\tilde{f}_{\lambda} ({\mathbf n})$. 

We now make the assumption, that the behaviour of the analyzers is
independent of context.

Let $\tau$ be any orthonormal triad which
$\tilde{f}_{\lambda}$-evaluates to one of the ``illegal''
combinations $000,011,101,110,111$.   We will  show that, if
$\tau$ is measured in the hidden state $\lambda$,   there is
probability
$\ge 0.5$ that the result  will be ``illegal''.

In fact, suppose that the $\tilde{f}_{\lambda}$-values of $\tau$
are $000$.  Let $p_r$ be the probability that the result of
measuring ${\mathbf e}_r$ is 0.  The defintion of
$\tilde{f}_{\lambda}$ implies that $p_r\ge 0.5 $ for each $r$.
The assumption of context-independence implies that the
probabilities are independent.  Consequently, the probability that
the measurement outcome will be one of the  ``legal''
combinations $001, 010, 100$ is
\begin{equation}
p_{\rm legal} =  p_1 p_2 (1-p_3) + p_1 p_3 (1-p_2)  + p_2 p_3
(1-p_1)
\end{equation}
It is  straightforward to verify that the expression on the right
hand side is always
$\le 0.5$.  A similar argument shows that $p_{\rm legal}$ is also
$ \le 0.5$ if the $\tilde{f}_{\lambda}$-values of $\tau$
are one of the other four ``illegal'' combinations.  We conclude
that, in every case, there is probability
$\ge 0.5$ that the measurement outcome will be ``illegal''---as
claimed.

Now let $\tau_1, \dots, \tau_N$ be a set of  triads which
cannot consistently be coloured.   For each $\lambda$ choose an
index $r_\lambda$ such that the 
$\tilde{f}_\lambda$-values of 
$\tau_{r_{\lambda}}$ are ``illegal''.   Let $\Lambda$ be the set
of all hidden states $\lambda$.  For each index
$r$ define
$\Lambda_r = \{ \lambda \in \Lambda \colon r_\lambda = r\}$. 
   On the one hand, it follows from the result proved in the last
paragraph that, if
$\lambda \in \Lambda_r$, then there is probability $\ge 0.5$ that
a measurement of $\tau_r$ will have an ``illegal'' outcome. On the
other hand the sets
$\Lambda_1,
\dots,
\Lambda_N$ partition
$\Lambda$ into
$N$ disjoint subsets, so there must be an index $r_0$
such that there is probability $\ge 1/N$ that $\lambda\in
\Lambda_{r_0}$.   Combining these statements we deduce that
there is probability $\ge 1/(2 N)$ that a measurement of
$\tau_{r_0}$ will have an ``illegal'' outcome.  However, it
follows from the definition  that this
probability must be
$\le
\epsilon_{\rm q}$.

We deduce that, if the behaviour of the analyzers is
context-independent, then $N \epsilon_{\rm q} \ge 1/2$.

\section{Sequential Measurements are Joint Measurements}
\label{ap:KentObjection}
Kent~\cite{MKC4}
has argued that a sequential measurement, such as the one
illustrated in Fig.~\ref{fig:sequential},
is not
a joint measurement.  According to him a process does not count as
a joint measurement unless the different observables are measured
all together, in parallel so to speak. The purpose of this
appendix is to explain why we consider this position to be
unjustified.

Measuring instruments
are  defined functionally, in terms of what they do.  This
is true both  classically and quantum mechanically.    If a
system performs the \emph{function}, 
of an instrument which measures
$\hat{A}$, then it \emph{is} an instrument which measures
$\hat{A}$. 

The function of a classical measuring instrument is to reveal 
pre-existing values.   Suppose that
one has  a sealed box with a dial and two terminals on the front,
and suppose that one wishes to confirm that  it is a classical 
ammeter.  This question can be decided without looking inside the
box.  It is sufficient to test the instrument with known currents,
and check that it reads correctly.  Anything which does the job of
a classical ammeter, is a classical ammeter.

A micrometer screw guage and an interferometer look very
different.  However, they both count as length measuring
instruments.  This is because they both do the job of a length
measuring instrument.

Similar considerations apply  in quantum mechanics. 
The function of a quantum measuring instrument is to discriminate
eigenstates.  Suppose one has a sealed box, with a hole into which
spin 1 particles can be injected, and a digital readout.  Suppose
it is found, after testing, that the box 
correctly identifies the joint eigenstates of
$({\mathbf e}_1 \cdot \hat{\mathbf S})^2$, 
 $({\mathbf e}_2 \cdot \hat{\mathbf S})^2$, 
 $({\mathbf e}_3 \cdot \hat{\mathbf S})^2$.  Then that is
enough to establish that it jointly measures these
observables.

Anything which discriminates the joint eigenstates of
$({\mathbf e}_1 \cdot \hat{\mathbf S})^2$, 
 $({\mathbf e}_2 \cdot \hat{\mathbf S})^2$, 
 $({\mathbf e}_3 \cdot \hat{\mathbf S})^2$, jointly measures these
observables.  In particular, the  apparatus illustrated in
Fig.~\ref{fig:sequential} jointly measures them.

A quantum measurement  can always be conceived in
``black-box'' terms.   In order to decide whether a process is a
quantum measurement of a given set of commuting observables it is
only necessary to consider the state of system+apparatus at the
time when the measurement interaction commences and again, at the
time when the interaction is concluded (the difference between
these times necessarily being $>0$). If the initial and final
states are related in the required manner, then the process is a
measurement of the observables concerned.   Questions as to the
detailed behaviour of the state between these times,  during the
course of the interaction, are irrelevant.


\begin{thebibliography}{99}
\bibitem{NotePred}
We are here focussing on what we have elsewhere~\cite{me3,me4}
described as the retrodictive aspect of a quantum measurement,
which is the only aspect relevant to the argument of this paper. 
For a more general discussion of quantum measurements, which takes
into account both the retrodictive and predictive aspects, see
Appleby~\cite{me3,me4}.
\bibitem{NotePOVM} We are indebted to Kent for clarifying this
point (in a personal communication), which we had not appreciated
at the time we wrote ref~\cite{me3}.
\bibitem{NoteIdealB} We have elsewhere~\cite{me3,me4} described~1
and~2 as the properties of retrodictive and predictive ideality
respectively.
\end{thebibliography}
\end{document}